\documentclass[a4paper,fleqn,usenatbib,useAMS]{mnras}

\usepackage{graphicx}	
\usepackage{amsmath}	
\usepackage{amssymb}	
\usepackage{multicol}        
\usepackage{bm}		
\usepackage{pdflscape}	
\usepackage{epsf}

\newcommand{\kms}{\;{\rm km}\,{\rm s}^{-1}}
\newcommand{\kmsmpc}{\kms\;{\rm Mpc}^{-1}}

\newcommand{\HI}{{\sc Hi}}

\newcommand{\hmpc}{h^{-1}{\rm Mpc}}

\newcommand{\msolar}{\;{\rm M}_{\odot}}

\newcommand{\giz}{{\sc Gizmo}}
\newcommand{\muf}{{\sc Mufasa}}

\usepackage[T1]{fontenc}
\usepackage{ae,aecompl}

\title[\muf: Galaxy conformity]{\muf: The strength and evolution of galaxy conformity in various tracers}

\author[Rafieferantsoa \& Dav\'e]{Mika Rafieferantsoa$^{1,2,3}$
\thanks{Contact e-mail: \href{mailto:rafieferantsoamika@gmail.com}
{rafieferantsoamika@gmail.com}}\thanks{South African
Astronomical Observatory, Observatory Road, Cape Town 7925, South Africa},
Romeel Dav\'e$^{5,1,3,4}$
\\
\\$^1$ University of the Western Cape, Bellville, Cape Town 7535,
South Africa
\\$^2$ Max-Planck-Instit\"ut f\"ur  Astrophysik, Garching, Germany
\\$^3$ South African Astronomical Observatory, Observatory,
Cape Town 7925, South Africa
\\$^4$ African Institute for Mathematical Sciences, Muizenberg,
Cape Town 7945, South Africa
\\$^5$ Institute for Astronomy, Royal Observatory, Edinburgh EH9 3HJ, UK
}

\date{Last updated 2017 December 18; in original form 2017 July 6}

\pubyear{2017}

\begin{document}
\label{firstpage}
\pagerange{\pageref{firstpage}--\pageref{lastpage}}
\maketitle

\begin{abstract}
We investigate galaxy conformity using the \muf\ cosmological
hydrodynamical simulation.
We show a bimodal
distribution in galaxy colour with radius, albeit with too many low-mass
quenched satellite galaxies compared to observations.
\muf produces conformity in observed properties such as colour,
sSFR, and \HI\ content; i.e neighbouring galaxies have similar properties.
We see analogous trends in other properties such as
in environment, stellar age, H$_2$ content, and metallicity.
We introduce quantifying conformity using ${\cal S}(R)$,
measuring the relative difference in upper and lower quartile properties of
the neighbours.
We show that low-mass and non-quenched haloes have weak conformity
(${\cal S}(R)\la 0.5$) extending to large projected radii $R$ in all properties, 
while high-mass and quenched haloes have strong conformity
(${\cal S}(R)\sim 1$) that diminishes rapidly with $R$ and disappears
at $R\ga 1$~Mpc. 
${\cal S}(R)$ is strongest for environment in low-mass haloes, and sSFR
(or colour) in high-mass haloes, and is dominated by one-halo conformity
with the exception of \HI\ in small haloes.  Metallicity shows a
curious anti-conformity in massive haloes.  Tracking the evolution
of conformity for $z=0$ galaxies back in time shows that
conformity broadly emerges as a late-time ($z\la 1$) phenomenon. 
However, for fixed halo mass bins, conformity is fairly constant with redshift
out to $z\ga 2$.  These trends are consistent with the idea that
strong conformity only emerges once haloes grow above \muf's quenching
mass scale of $\sim 10^{12}M_\odot$.  A quantitative measure of
conformity in various properties, along with its evolution, thus
represents a new and stringent test of the impact of quenching
on environment within current galaxy formation models.
\end{abstract}




\section{Introduction}

\label{intro}
\indent

Environment plays an important role in setting the properties of
galaxies.  The collapse of massive haloes and large filaments results
in the shock heating of gas, which inhibits the growth of galaxies
within these structures by a combination of strangulation (or
starvation), ram pressure stripping, and tidal stripping.  These
environmental processes connect the physics of intergalactic gas
and large-scale structure with the observable stellar properties
of galaxies, and hence represent a key test for cosmologically-situated
galaxy formation models.

A particular phenomenon that has gained attention recently is the
tendency for galaxies spatially close to each other to have similar
galaxy colours.  This was first noted in \citet{Weinmann-06-a} and
given the name {\it galactic conformity}.  They used a galaxy sample
from the Sloan Digital Sky Survey \citep[SDSS]{York-00} and found
that early-type central galaxies have comparatively higher fractions
of early-type satellite galaxies around them, while late-type
centrals tended to be surrounded by late-type satellites. 
Furthermore, the overall fractions of early and late types
depended strongly on the mass of their respective haloes. 
This type of conformity was later dubbed ``one-halo" conformity,
since it quantifies the level of similarity within a single halo.

Conformity can arise owing purely to large-scale structure heating,
because of the tendency of deep potential wells to shock-heat gas
to high temperatures~\citep[e.g.][]{White-78}.  Such shock-heated
gas is expected to surround galaxies in haloes with masses above
$\sim 10^{12}M_\odot$~\citep{Birnboim-Dekel-03,Keres-05,Gabor-10}, and is
likely related to the oft-mentioned bimodal distribution in galaxies
properties \citep[e.g.][]{Kauffmann-03, Baldry-04}.  Such bimodality
is present in both central and satellite galaxy samples.  This
encompassing hot halo can thus naturally give rise to a correspondence
between central and satellite star formation rates and gas contents.

But such a hot halo is not stable.  The hot gas in the dense central
region is expected to cool quickly, leading to substantially star
formation in massive galaxies, in disagreement with observations.
Thus most galaxy formation models introduce some feedback
mechanism~\citep{Somerville-15r}, putatively from the central active
galactic nucleus (AGN), that maintains the hot hydrostatic
halo~\citep{Croton-06}.  This same ``maintenance mode" feedback can
attenuate the star formation in satellite galaxies as well
\citep[e.g.][]{Ann-08,Gabor-Dave-12}.  Hence one-halo conformity may
encode information both about large-scale structure as well as AGN
feedback processes in more massive systems.

\citet{Kauffmann-13} re-analysed galactic conformity in SDSS with
a more rigorous sample selection, and found that moderate-mass
central galaxies show galactic conformity out to distances that are
many times their virial radius, as much as 4~Mpc, which has come
to be called two-halo conformity since it corresponds to properties
being similar in galaxies living in different haloes.  In contrast,
more massive centrals only show conformity to the neighbours within about
a virial radius.  To explore the physical origin of conformity,
\citet{Kauffmann-15} further found that central galaxies with low
star formation rate are more likely to be located in a neighbourhood
with higher fraction of massive galaxies that have active AGN,
suggesting that AGN feedback plays a significant role in conformity.

However, the strength or perhaps even the existence of two-halo conformity
is controversial.  Recent work by \citet{Tinker-17} suggested that
the large spatial extent of galactic conformity found in
\citet{Kauffmann-13} may be the result of misclassifications of 
central galaxies, and that conformity beyond at most a Mpc disappears
when the central sample is more carefully selected using a group
catalog rather than projected distance.  This is supported by
\citet{Sin-17}, who further showed that the strong two-halo conformity
signal found by \citet{Kauffmann-13} can be reproduced with their
semi-analytic model when using Kauffmann et al.'s selection, but
when true central galaxies are used, the two-halo term is very
weak and essentially undetectable in the SDSS data.

Conformity analogously appears in the neutral hydrogen content of
galaxies.  Using 40 galaxies from the Bluedisk project, \citet{Wang-15}
found that galaxies with high \HI~fraction live in the vicinity of
other galaxies with high \HI~fraction.  Also, conformity persists
out to higher redshift.  \citet{Hartley-15} studied the satellite
galaxies drawn from the UKIRT Infrared Deep Sky Survey
\citep[UKIDSS][]{Lawrence-07} and concluded that passive galaxies
are more likely to be around passive galaxies with $3\sigma$
significance and that happens out to $z\ga 2$.  Using a
set of satellite galaxies data from ZFOURGE \citep{Straatman-14},
UDS, and UltraVISTA~\citep{MacCracken-12}, \citet{Kawinwanichakij-16}
looked at the evolution of galactic conformity and confirmed a
one-halo conformity signal with a significance of more than $3\sigma$
out to $z\sim1.6$, and a lower but noticeable signal out to $z\sim2.5$.
Similarly, \citet{Berti-16} investigated one- and two-halo conformity
with the Primus survey \citep[conducted with IMACS;][] {Bigelow-03}
data which they claimed to be uniquely suited due to large survey
area of $\sim9$ deg$^2$ and redshift precision of $\sigma_z=0.005(1+z)$.
They detect more than $2.5\sigma$ one-halo conformity out to $z\sim1$.
They also found a hint at a two-halo conformity signal from the
fact that central galaxies are more likely to be quiescent when
they are located in dense environment.  \citet{Hatfield-16} used a
different method by looking at the cross-clustering of galaxies
with the 2-point correlation function and claim that specific star
formation rate (sSFR)-density relation, which is another way of
characterising conformity, emerges at $z\sim1$ and keeps growing
until today.  Hence, it appears that conformity is a real effect not only
in present-day galaxy colours, but in gas content at least, as well
as to higher redshifts, though the precise strength and evolution 
depends on the sample selection and technique used to quantify 
conformity.

Given the emergence of this wealth of data on galaxy conformity,
there has been various attempts to explain conformity within a
hierarchical structure formation paradigm.  \citet{Hearin-15} used
semi-analytic models on the Bolshoi simulation \citep{Klypin-11}
and found that using either M$_{\mathrm{halo},vir}$-based quenching
prescription or a delayed-then-rapid quenching of the galaxies
displayed zero conformity, while only their {\it age matching} model
showed statistically significant galactic conformity.  This led
them to conclude that two-halo conformity is the result of the central
galaxy assembly bias.  \citet{Hearin-16} followed up by looking at
the assembly histories of structures and found that haloes separated
with more than tens of their virial radius are connected because
they are situated within the same large-scale tidal environment
which is the main driver of their growth.  However, \citet{Zu-Mandelbaum-17}
used a colour-based halo occupancy model to argue that (weak)
large-scale conformity can arise purely from the environmental
dependence of the halo mass function, without requiring any assembly
bias.  They also confirm the result of \citet{Sin-17} and
\citet{Tinker-17} that the strong two-halo conformity seen by
\citet{Kauffmann-13} is primarily an artifact of mis-identified
central vs. satellite galaxies.

Modern cosmological hydrodynamic simulations include all the relevant
effects that are expected to give rise to galactic conformity.  Such
models should, in principle, implicitly include halo assembly bias,
halo occupancy evolution, and any correlations between the colours
of centrals and satellites arising from included feedback processes,
and hence galaxy conformity should be an emergent property.  Such
simulations track gas and star formation properties directly as
well, so can be used to study conformity in various tracers, as
well as their evolution with redshift.  Nonetheless, as shown in
e.g. \citet{Gabor-Dave-15}, the quenching of satellites along with
surrounding ``backsplash" and ``neighbourhood quenched" galaxies is
dependent on the model for central galaxy quenching, which is at
present not well constrained in galaxy formation models, and is
typically included only in a prescriptive or heuristic manner.
Hence it is instructive to examine conformity predictions from
cosmological hydrodynamic simulations, and compare them to present
and future observables as a detailed test of quenching models.

To this end, \citet{Bray-16} used the Illustris
Simulation~\citep{Vogelsberger-14} to explore whether assembly bias
can be an explanation for galaxy conformity.  They found evidence
of galactic conformity out to 10 Mpc for the smallest centrals,
decreasing in strength with increasing stellar mass of the centrals.
They developed a simple model based on abundance and age matching
that was able to reproduce this signal, demonstrating that the
galaxy colour-age relation is important for conformity.

In this paper, we use the \muf\ simulation~\citep{Dave-16} to study
conformity.  Our goal is to investigate conformity in a wide variety
of galaxy tracers, to understand which galaxy and environmental
parameters show the strongest levels of conformity, and to make
predictions for conformity that can be used as a test of models.
This differs in aim from previous works discussed above that have
focused more on developing halo-based models for the origin of
conformity, comparing to the (sparse) available data.  We show that
\muf\ predicts a conformity signal that is strongly dependent on
halo mass, and that this signal is not limited to colour and neutral
hydrogen but appears in many other galaxy properties as well.  We
propose a measure to quantify the strength of conformity and study
its evolution in various tracers as a function of redshift.  Our
results indicate that galaxy conformity is a generic emergent feature
of hydrodynamic galaxy formation models, and that the strength of
conformity can be a valuable test of the interplay between environment,
galaxy assembly, and feedback processes particularly related to
quenching.

\S\ref{sim} briefly reviews the \muf\ simulation used for this work.
In \S\ref{gal_sat_prop}, we look at the satellite galaxy properties
from our simulation. \S\ref{gal_conf} expands on the comparison of
conformity between our simulated galaxy sample and the observed
data. \S\ref{gal_conf_ev} characterises the nature of conformity
comparing between various tracers, and study its evolution out to
intermediate redshifts.  We summarize our conclusions in
\S\ref{conclusion}.

\section{Simulations}\label{sim}
\subsection{Models}\label{model}
\indent

For this work we employ the \muf\ simulation, which is fully described
in \citet{Dave-16}.  Here, we briefly review the main ingredients,
and expound on the key physical aspects of \muf\ that are particularly
relevant for this work.

\muf\ employs the \giz\ cosmological hydrodynamic code, including
a tree-particle-mesh gravity code based on {\sc
Gadget}~\citep{Springel-05}, and a meshless finite mass hydrodynamic
algorithm \citep{Hopkins-15}.  For radiative processes, \muf\
utilises the {\sc Grackle 2.1}
library\footnote{\url{https://grackle.readthedocs.io/en/grackle-2.1/genindex.html}}
to cool the gas elements, accounting for non-equilibrium ionisation
for primordial elements, as well as metal-line cooling assuming
ionisation equilibrium,  plus photoionisation heating computed
with a spatially-uniform metagalactic flux
taken from \citet{Faucher-Giguere-10}.
Star formation occurs in molecular gas, and only in 
gas elements with
hydrogen number density $n_H\geq0.13\ \mathrm{cm^{-3}}$,
with the star formation rate computed following a 
\citet{Schmidt-59}-law scaling, namely 
\begin{equation}\label{sl}
{\rm SFR} = \varepsilon f_\mathrm{H_2} G^{-0.5} \rho_\mathrm{gas}^{0.5}.
\end{equation} 
Here $\rho_\mathrm{gas}$ is the density of the gas,
$\varepsilon=0.02$ \citep{Kennicutt-98} is the star formation
efficiency, $G$ the gravitational constant, and $f_\mathrm{H_2}$
the molecular hydrogen fraction in the gas element computed via
a subgrid prescription from \citet{Krumholz-Gnedin-11}.

We assume that star formation in the simulation produces a combination
of radiation pressure and supernovae energy which manifests by
kicking out its surrounding gas volume elements at a given rate
$\eta$ relative to the star formation rate.  Each outflowing wind
element is ejected away from its host galaxy and in a direction
perpendicular to the ($\vec{\textbf v}$, $\vec{\textbf a}$) plane with a
launching wind velocity $v_w$, where $\vec{\textbf v}$ and $\vec{\textbf
a}$ are the velocity and the acceleration of the gas cloud prior
to its launch. To choose the free parameters, we take scaling
relations from \citet{Muratov-15} based on the Feedback in Realistic
Environments (FIRE) simulations.  In particular, we choose the mass
loading factor $\eta$ and the wind speed $v_w$ as follows:
 \begin{equation}\label{wind}
 \eta = 3.55\left(\frac{M_*}{10^{10}\msolar}\right)^{-0.35};
 \end{equation}
 \begin{equation}\label{windvel}
v_w = 2 v_c \left(\frac{v_c}{200\kms}\right)^{0.12}.
\end{equation}
$M_*$ and $v_c$ are the stellar mass and circular velocity of the
galaxy where the gas volume element is located.  Galaxies are
identified using an approximate on-the-fly friends-of-friends group
finder specifically designed to be computationally fast and tuned
to reproduce the same results as {\sc
Skid}\footnote{\url{http://www-hpcc.astro.washington.edu/tools/skid.html}\\
{Our sample from skid does not contain dark matter depleted galaxy
(see Figure~ \ref{dm_content}).}
\label{skid}}
(Spline Kernel Interpolative Denmax); this is applied only to star-forming
gas elements and stars.

Particularly relevant for this paper is that \muf\ includes an
observationally motivated heuristic prescription to quench massive galaxies.
Gas elements sitting in a host halo above a threshold quenching
mass $M_q$ are heated to around the virial temperature of that host.
This is done except for the interstellar medium gas defined to have
more than 10\% neutral hydrogen fraction.  The host halo is grouped
on the fly with a (separate) friends-of-friends algorithm using a
linking length of 0.16 times the mean inter-particle distance,
including dark matter, gas, and stars.  The virial temperature is
taken to be $T_{vir} = 9.52\times 10^7 M_h^{2/3}$ \citep{Voit-05}.
$M_q$ is taken to be redshift dependent, whose scaling is taken
from the analytic equilibrium model of galaxy formation \citep{Mitra-15},
who obtained a best-fit scaling of
$M_q = (0.96+0.48z)\times10^{12}\msolar$.

We note that this quenching prescription is purely heuristic, and
is not a physical model for AGN feedback.  It is specifically
designed to mimic the effects of radio mode feedback~\citep{Croton-06}
from active galactic nuclei (AGN), in which jets from the central
galaxies of massive haloes are observed to add enough energy into
diffuse gas to counterbalance cooling~\citep{McNamara-07}, without
a detailed physical model for the AGN energy couples to the halo
gas.  Since we keep ambient gas hot all the way out to the virial
radius in massive haloes, this feedback model can be regarded as a
rather extreme form of maintenance mode quenching.  In \citet{Dave-17b}
we have showed that it results in a population of quenched central
galaxies that is in reasonable agreement with key observations such
as the colour-magnitude diagram of galaxies, although it appears
to overproduce quenched satellite galaxies at low masses.  The
results on conformity here should be taken with these caveats in
mind, noting that a different or more physical quenching model may
yield different results.

\subsection{Galaxy sample and operational definitions}\label{simg}

The galaxy sample used for our analysis is obtained by simulating
a cube of $50\hmpc$ on a side with $512^3$ dark matter particles
and $512^3$ gas volume elements. The initial conditions are generated
at redshift $z=249$ using {\sc Music} \citep{Hahn-11} with
\citet{-16}-concordant cosmological parameters, namely $\Omega_m =
0.3$, $\Omega_\Lambda = 0.7$, $\Omega_b = 0.048$, $H_0 = 68 \kmsmpc$,
$\sigma_8 = 0.82$ and $n_s = 0.97$.  We also consider a $25\hmpc$
volume with the same number of particles and the same input physics
and cosmology, having a factor of 8 better in mass resolution, in
order to assess resolution convergence.

\muf\ evolves these initial conditions to $z=0$ outputting 135
snapshots.  For each snapshot, we identify galaxies, with {\sc
Skid}\textsuperscript{\ref{skid}}~\citep{Keres-05}, as gravitationally
bound collections of stars and star-forming gas. For galaxy
mass resolution, we take $5.8\times10^8 \mathrm{M_\odot}$,
where it was shown that the stellar mass functions converge with that limit
\citep{Dave-16}.  While this does
not guarantee convergence in other properties such as colour, we
will compute conformity using galaxies with $M_*>1.8\times 10^9M_\odot$,
three times larger than our nominal mass resolution and corresponding
to approximately 100 star particles at minimum. Host haloes are
identified using the friends-of-friends algorithm with a linking
length of 2\% the mean inter-particle distance. The galaxies are
assigned to host haloes based on their spatial location. The most
(stellar) massive galaxy in a given halo is considered as the central
(regardless of physical location) and the remainder are satellites.

\begin{figure}
\includegraphics[scale=1.0]{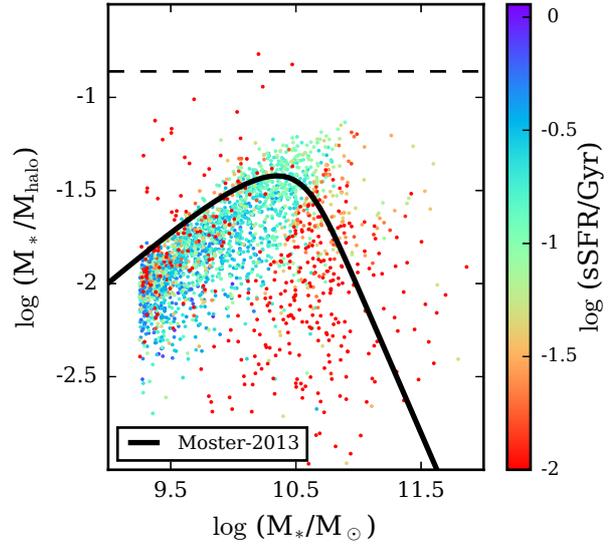}
\vskip-0.2in
\caption{Stellar-to-halo mass ratio as a function of $M_*$
for central galaxies in \muf.  The ratio peaks around $M_*\sim 10^{10.5}M_\odot$
at around 25\% of the cosmic baryon fraction (shown as the horizontal dotted line),
dropping away quickly to higher and lower masses.
The solid line shows the fit from abundance matching of observed galaxies
by \citet{Moster-13}; the predicted values are generally consistent with this.
}
\label{dm_content}
\end{figure}

Figure~\ref{dm_content} shows the stellar-to-halo mass ratio
in central galaxies as a function of stellar mass, colour-coded by
specific SFR.  This shows that quenching kicks in rather rapidly
at $M_*\ga 10^{10.5}M_\odot$, above which the typical halo mass
lies in the regime where \muf's halo quenching model kicks in.
There are a handful of galaxies with high $M_*/M_{\rm halo}$ ratios
at lower masses, which tend to be quenched; these are typically
former satellites whose orbits have taken them outside their host
haloes~\citep{Gabor-Dave-15}, and thus have had their own haloes significantly
impacted by stripping.  Such galaxies should also appears in
observational samples, and indeed there is strong increase in the
passive fraction of galaxies lying close to but still outside the
virial radius of massive galaxies~\citep{Geha-12}.  These have only
a small impact on the overall conformity statistics.

The properties of these galaxies and haloes are calculated with a
modified version of {\sc
caesar}\footnote{\url{https://bitbucket.org/laskalam/caesar}}, which
is an add-on package for the {\sc y}t simulation analysis suite.
The stellar mass (M$_*$) of a galaxy is obtained by summing the
masses of all its stellar particles, and we define the galaxy's age
as the time when half of these stars were formed.  The galaxy star
formation rate (SFR) is the summation of the instantaneous SFR of
the gas elements (directly obtained from the simulation).  The molecular
hydrogen fraction ($f_\mathrm{H2}$) is the total mass of H$_2$ from
all the gas elements, which is tracked directly in the simulation,
divided by its stellar mass.  The atomic hydrogen (\HI) content of
the galaxy is the aggregate amount of all \HI\ from the gas particles.
Before summing over, the gas \HI\ mass from the simulation is
post-processed to account for the self-shielding from the metagalactic
UV background radiation, by using the fitting formula for the
effective optically-thin photoionization rate as a function of
density taken from \citet[][see eq.  1]{Rahmati-14}. The \HI\
richness ($f_\mathrm{HI}$, or \HI\ fraction) is the total \HI\
content of the galaxy divided by its stellar mass M$_*$.   
To quantify the environment,
we use the projected nearest neighbour density $\Sigma_3$:
\begin{equation}
\Sigma_3 = {3\over \pi R_3^2}
\end{equation}
where $R_3$ is the distance of the galaxy to its 3rd closest 
neighbour, projected along the z-axis.

The colours of the galaxies are obtained using the {\sc
Loser}\footnote{Line Of Sight Extinction by Ray-tracing\\
\url{https://bitbucket.org/romeeld/closer}} package.  Stellar spectra
are interpolated from the age and the metallicity of star particles
using the Flexible Stellar Population Synthesis \citep[FSPS;][]{Conroy-10}
library. The metal column density is calculated along each line of
sight, converted into a dust extinction, then applied to each star
particle's spectrum.  The spectra of all the stars in each galaxy
(from {\sc skid}) are then summed and the appropriate filter applied
to get the magnitudes.  See \citet{Dave-17b} for further details.

To analyse conformity, we follow the general procedure outlined in
\citet{Kauffmann-13}.  First we subdivide our $z=0$ central galaxies
into three stellar mass bins: $\log (M_{*,\mathrm{cen}}/\msolar)
\in \{[9.5, 10], [10, 10.5], [10.5, 11.5]\}$. The numbers of
galaxies in each bins are 747, 587, and 561, respectively.
We choose $10.5-11.5$ for the largest bin (instead of $11-11.5$ that was used in
\citealt{Kauffmann-13}) because using smaller bin results in
considerable shot noise owing to small numbers of such galaxies in
our simulation; we checked that the results for our larger bin are
consistent with that obtained from using only $M_*>10^{11}M_\odot$,
as both predominantly live in quenched haloes.  

Within each stellar mass bin, we order the central (or primary)
galaxies by a given property: colour,
sSFR or \HI\ richness (M$_\mathrm{{\sc HI}}/M_*$).  We then take
the objects at the lowest and the highest quartiles ($<25\%$ and
$>75\%$), and examine the median properties of neighbours of these
galaxies as a function of radius out to 4 projected Mpc.
Throughout our analysis, we employ jackknife resampling among 8
simulation sub-octants to estimate our errors.

Guided by the completeness level in the SDSS-based sample of
\citet{Kauffmann-13}, we only consider satellite galaxies with
$M_*\geq10^{9.25}M_\odot$, comfortably above our stellar mass
resolution limit.  However, unlike \citet{Kauffmann-13}, we use
friends-of-friends identified central galaxies instead of adopting
their isolation criteria.  This is more closely aligned with more
recent analyses that use group catalogs, which has been shown to
provide a more robust measure of conformity, particularly two-halo
conformity, compared to the Kauffmann et al. isolation criterion
\citep{Tinker-17}.  

\begin{figure}
\includegraphics[scale=0.8]{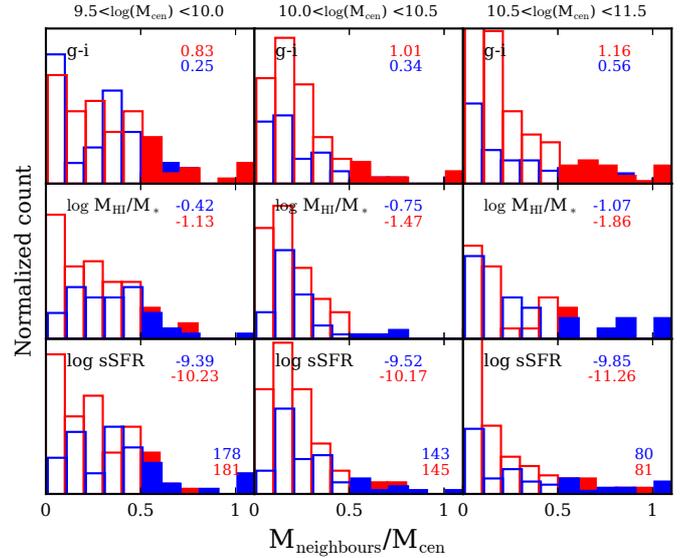}
\caption{Stellar mass fraction of the neighbours relative to their respective
central galaxies in our simulated box. We only show galaxies that are within
500 kpc projected distance and 500 $\kms$ redshift distance from
their primaries. The histograms at $>0.5$ are filled to emphasize on 
the fractions of neighbours where their respective central galaxies would
not have been classified as isolated galaxies if Kauffmann criteria were applied.}
\label{mass_ratio}
\end{figure}

Figure \ref{mass_ratio} illustrates the difference in mass ratio
of identified centrals to their neighbouring galaxies located
within 500 projected kpc and 500 $\kms$ redshift distance,
which corresponds to the Kauffmann isolation criterion.
The distribution of galaxies in the upper and lower quartiles
in galaxy colour (top row), \HI\ fraction (middle),
and sSFR (bottom) are indicated in red and blue histograms, in
three central mass bins (left to right columns). 
The numbers on the bottom right show
the number of central galaxies in the respective stellar mass bin.
This shows that overall a small fraction
of the neighbours have M$_*> 0.5\times\mathrm{M_{primary}}$
(filled part of the histograms)
where their respective central galaxies would not have been classified as isolated
with respect to Kauffmann criteria.  Given the relatively modest fraction of 
such neighbours ($\sim 15\%$) our conformity should be
a robust prediction independent of isolation criterion, but for
high-mass systems there will be systematic differences.
Note that if we only used the satellite galaxies (not shown),
the fraction of neighbours with M$_*> 0.5\times\mathrm{M_{primary}}$
drops significantly to $\sim 5\%$.
We will therefore compare to the \citet{Kauffmann-13} data for illustrative
purposes only, with the caveat that the isolation criterion can play
some role in the outcome.

\section{Satellite galaxy properties in \muf} \label{gal_sat_prop}

Conformity is a measure of the environmental impact on galaxy
formation.  Hence it is important to ensure that environmental
processes are reasonably accurately modeled in our simulation.  As
a test of this, we begin by presenting an analysis of the satellite
galaxy population predicted in \muf.

\subsection{Satellite versus central mass functions}

\begin{figure*}
\includegraphics{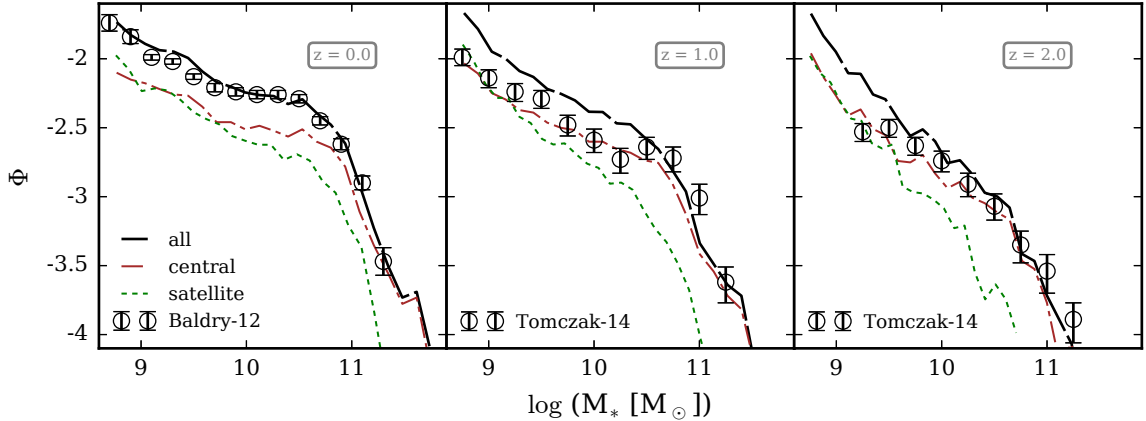}
\caption{Galaxy stellar mass functions (black long-dashed lines)
separated into central (brown dashed-dotted lines)
and satellite (green dotted lines), at redshift $z=0$
(left), $z=1$ (middle) and $z=2$ (right). At all epochs, 
central galaxies dominate by number at $M_*\ga 10^{9-9.5}M_\odot$,
and a knee begins to appear in the satellite mass function at
$z\la 1$. Observations of the total GSMF are shown with the black circles.
}
\label{fig_gsmf_cen_sat}
\end{figure*}

Figure \ref{fig_gsmf_cen_sat} shows the GSMF of the \muf\ simulated
galaxies (black long-dashed lines): separated into central (brown
dashed-dotted lines) and satellite (green dotted lines) galaxies at
$z=\{0, 1, 2\}$. Observational data from \citet{Baldry-12} for $z=0$
and \citet{Tomczak-14} for $z=\{1,2\}$ are shown with the black
circles. 

As discussed in \citet{Dave-16}, \muf\ reproduces the observed
total GSMF and its evolution out to high redshifts fairly well,
albeit with a modest excess at $z\sim 1$.  The massive end of the
GSMF consists almost entirely of central galaxies across all redshift
explored here. The contribution from the satellites is present only
at the low mass end ($M_*\la 10^{9.5}M_\odot$) and it is modestly
stronger at lower redshift.

Already at $z=2$, the central GSMF starts to produce the knee while
the satellite galaxies only show a hint of a knee at $z\la 1$. The knee
in GSMF is the result of an enhancement in star formation due to
an increased contribution of wind recycling to higher
masses~\citep{Oppenheimer-10}, combined with the truncation of star
formation in massive galaxies owing to our quenching prescription
mimicking AGN feedback.  At later epochs, the most massive satellites
were until recently central galaxies that were massive enough to
experience quenching, and hence they reflect a truncated high-mass
GSMF of the centrals.

\subsection{Star-forming versus quiescent satellite mass functions}\label{sat_mf}

\begin{figure*}
\includegraphics{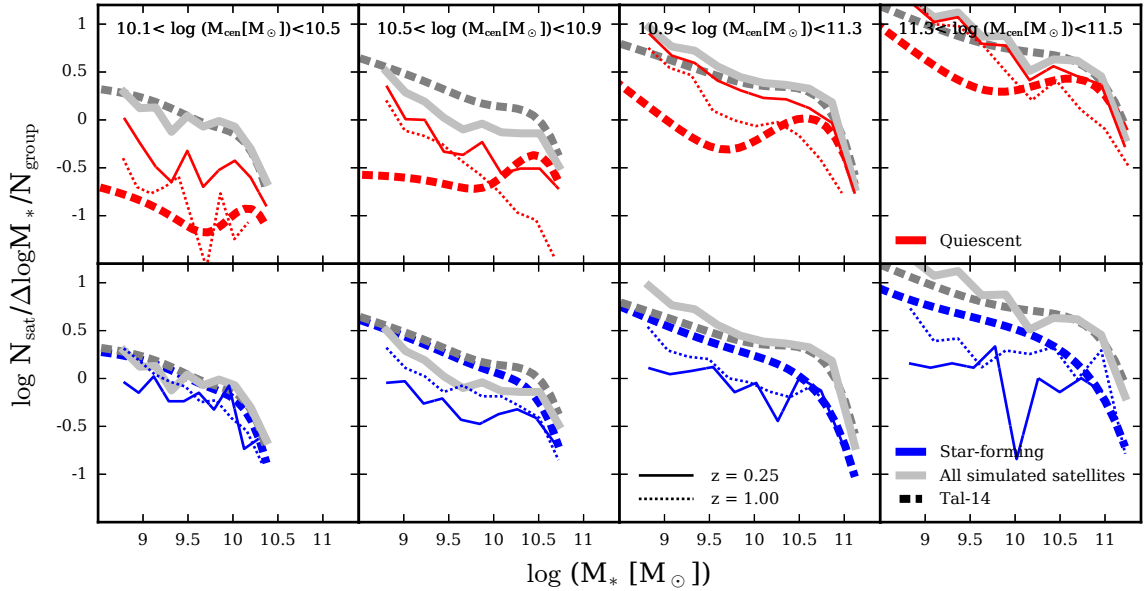}
\caption{Satellite galaxy stellar mass functions (thick grey lines, $z=0$);
separated into star forming (lower panels, blues) and
quenched (upper panels, reds) population. 
We show them at two different redshift $z=0.25$ (thin solid lines),
and $z=1$ (thin dotted lines). The thick dashed lines
show the double-Schechter fits from \citet[][see Table 1]{Tal-14}:
blue for star forming,
red for quenched and gray for all satellites.}
\label{fig_sat_mf}
\end{figure*}

Figure \ref{fig_sat_mf} breaks down the GSMF in terms of satellites
split into star-forming and quiescent within four central galaxy
stellar mass bins, following the same exercise as in \citet{Tal-14}
using UltraVista data.  Here we consider galaxies to be quiescent
if $\log (\mathrm{sSFR/Gyr}^{-1}) \leq -2.2$.  In accordance with
\citet{Tal-14}, we define $\Phi$ to be the number of galaxies per
group. In other words, for the sample of all the N central galaxies
within a given stellar mass bin, Figure~\ref{fig_sat_mf} shows the
distribution of their satellite galaxies normalized by N. In their
analysis, \citet{Tal-14} defined galaxies to be central if no other
more massive galaxy could be found within two projected virial
radii, which is broadly consistent with our definition to be the
most massive galaxies in their respective haloes (see sec. \ref{simg}).
They consider all non-central galaxies within two virial radii of
the central galaxies to be their satellites.

The total predicted satellite GSMF, as indicated by the thick solid
grey lines, are in general agreement with the \citet{Tal-14}
observational data shown by the thick dashed gray lines. There is
a clear trend of a higher mass function of satellites around more
massive centrals.  The only notable discrepancy is that the
observations show up to $\sim\times 2$ more satellites at intermediate
masses in the $10^{10.5}-10^{10.9}M_\odot$ central stellar mass
bin. At higher central mass bins, simulations tend to slightly have
more satellites than the observations at the lowest mass ends. Hence
overall, the total number of satellites is in broad agreement, with
a hint that \muf\ underproduces the satellite population at
intermediate masses ($\sim 10^{10}M_\odot$).

In contrast, we see more substantial discrepancies when examining
the satellite GSMFs broken down by quiescent vs. star-forming.  At
the lowest central masses, most satellites are blue (star-forming),
and \muf\ reproduces those well.  However, \muf\ strongly over-predicts
the (small) number of red satellites, particularly at the lowest
satellite masses.  At higher central masses, the observed red satellite mass
function is relatively shallow (thick red dashed lines), albeit
with an upturn at the lowest masses, whereas \muf\ predicts a steeper
red satellite GSMF.  Particularly, for the most massive centrals,
\muf\ predicts that red satellite galaxies dominate at all masses,
while in \citet{Tal-14} they only dominate at $M_*\ga 10^{10.5}M_\odot$.

The overall redshift evolution predicted in \muf\ is qualitatively
consistent with that seen by \citet{Tal-14}, where they used three
redshift bins such as $(0.2,0.5)$, $(0.5,0.85)$ and $(0.85, 1.20)$.
For illustration, we only show the simulated sample at $z=0.25$
(thin solid lines) and $z=1$ (dotted lines) which span the redshift
ranges used in their sample.  The star-forming satellite mass
functions show very little evolution with time, with a hint of being
slightly higher at higher redshift.  They dominate the mass function
at low masses, and are more prevalent at earlier epochs. Meanwhile,
there is a mild increase with time for the quenched satellite
population, but the trend with mass is much more significant; there
is a much larger number of red satellites per group around massive
centrals, and at high masses they dominate over the star-forming
satellites.  These general trends are qualitatively in agreement
with the observations.  However, as noted before, \muf\ strongly
over-produces the number of low-mass red satellites, at essentially
all central masses.  Moreover, \muf\ tends to grow massive red
satellites more rapidly than the low-mass red satellites, which is
opposite to what is seen in the data in which the most massive
satellites are already in place fairly early on.

Several avenues could lead to these discrepancies.  For instance,
\muf\ identifies central galaxies via a 3-D friends-of-friends
scheme, while in \citet{Tal-14}, centrals are identified as having
no other more massive system within two projected virial radii,
which can blend systems in projection.  In particular, it could be
that the ``bump" of massive red satellites identified by \citet{Tal-14}
are actually nearby massive central galaxies.  Furthermore,
\citet{Tal-14} must do substantial background subtraction to count
satellites, which can be uncertain particularly when low numbers
of satellites are present.  We could in principle mimic the first
selection effect, but the second effect would require detailed
modeling of the background population, which is beyond the scope
of this work.  In general, without a more careful mimicking of the
observational selection effects, the significance of these discrepancies
is not completely clear.

If we take the discrepancies at face value, it suggests that \muf\
over-quenches satellites in haloes of all masses, particularly low
masses satellites.  This is consistent with the findings in
\citet{Dave-17b}, but here we see this across a range of mass bins
and epochs.  The discrepancy for high-mass centrals may owe to the
extreme nature of our quenching mechanism, where we keep gas hot
all the way out to the virial radius.  For low-mass centrals,
however, the quenching mechanism is not obviously relevant, except
if such low-mass centrals happen to be in the vicinity of high-mass
centrals~\citep[``neighbourhood quenched";][]{Gabor-Dave-15}.  Hence
the discrepancies may reflect details of hydrodynamic processes,
or else some of the systematic effects mentioned in the previous
paragraph.  For now, we will consider the broad agreement as
sufficient for examining galaxy conformity in \muf, with the caveat
that \muf\ does not well reproduce the low-mass satellite
population when broken up by colour.

\subsection{Halo-centric satellite colours}

Conformity was first specified as a commonality in colours between
nearby centrals and satellites.  Hence an important aspect to
investigate in our models is the colours of satellite galaxies as
a function of radius.  In this section we examine the halo-centric
colours of satellite galaxies.

\begin{figure*}
\includegraphics{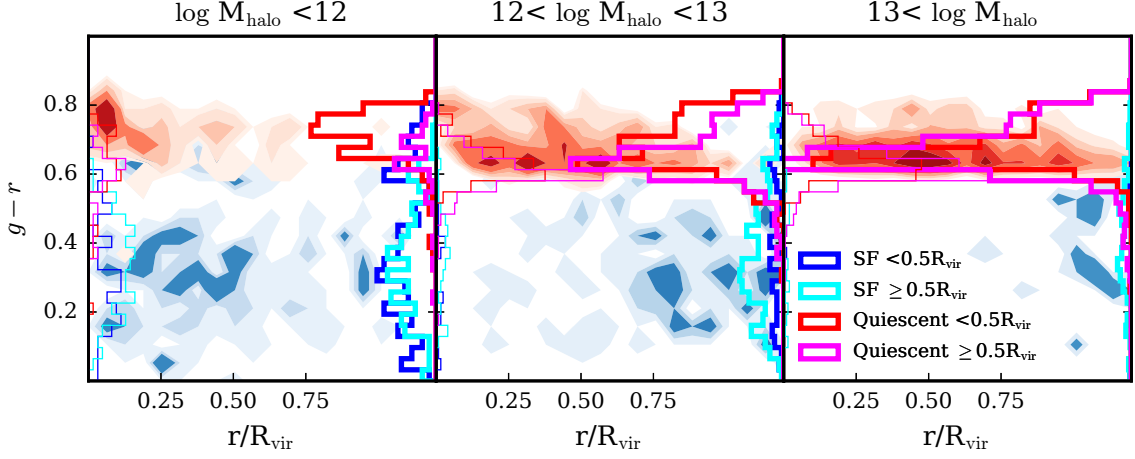}
\caption{$g-r$ colour of neighbouring galaxies depending on their distances
to the centre of their corresponding haloes ($z=0$).
The red (blue) shaded area shows the distribution
of the quiescent (star forming) satellites in our simulated sample.
The histograms along the right y-axes show the distribution of the satellites
depending on their specific star formation rate and their distance
to the halo centers. We also show the colour distributions of our smaller box
(higher resolution, but poorer statisctics) on the left y-axes.}
\label{fig_color_halo}
\end{figure*}

Figure \ref{fig_color_halo} shows $g-r$ contour plots of all
satellites in \muf, as a function of radius scaled by the virial
radius R$_{\rm vir}$.  The red shaded area shows the quiescent
satellites, while the star-forming satellites are depicted by the
blue shaded area; we remind the reader that these are divided at
sSFR=$10^{-2.2}$ Gyr$^{-1}$.  The right y-axis histograms show the
colour distributions of the satellites: blues (reds) for star forming
(quiescent) satellites at $r< 0.5$R$_\mathrm{vir}$, and cyan (magenta)
for star forming (quiescent) satellites at $r\geq 0.5$R$_\mathrm{vir}$.
We show for three different halo mass bins.  As a test of
numerical convergence, we also show the colour distributions of our
$25\hmpc$ box in thin histograms on the left y-axis.

We can see a strong bimodality distribution in terms of colour that
extends out to the virial radius.  This bimodality exists for all
halo masses, although the relative number of red and blue satellites
changes substantially with halo mass.  There are a few dusty
star-forming satellites that have red colours lying underneath the
red contour, but these are small fraction of the total.  The strong
bimodality in colour was first noted observationally by
\citet{Kauffmann-03,Balogh-04}, and is interpreted to indicate that
satellite galaxies quench fairly rapidly once the quenching process
begins~\citep[e.g.][]{Wetzel-15}.  \muf\ broadly reproduces these
observed trends.  The thin histograms from the $25\hmpc$ volume
along the left y-axis are qualitatively similar, 
although the bimodality is somewhat weaker maybe because the
smaller volume does not produce as many massive quenched systems
and the model performs differently at that resolution.

Examining the trends with halo size, we clearly see a decrease of
star forming population towards larger halo masses.  For less massive
haloes, we see that quiescent satellites are mostly located at the
core (inner $\sim 10-15$\%) of the groups while the star forming
satellites are almost evenly distributed. For intermediate-mass
haloes, the star forming satellites inside $<0.5\mathrm{R}_\mathrm{vir}$
rarify and the quiescent galaxies extend out to
$\geq0.5\mathrm{R}_\mathrm{vir}$.  For the most massive haloes,
only the satellite galaxies at the very edge of the haloes are
forming stars. These trends are again qualitatively consistent with
observations, as more massive central galaxies tend to be quenched
and have more environmentally-quenched satellites~\citep[e.g.][]{Peng-12}.

Overall, these results taken together generally indicate that the
satellite population in \muf\ qualitatively reproduces observations,
including the satellite population evolution split by red versus
blue galaxies.  The most notable discrepancy is that \muf\ predicts
an excess of low-mass red satellites in massive haloes.
Broadly, this is expected to strengthen the conformity of red
centrals with red neighbours in massive haloes, hence the predicted
conformity is likely to be overestimated in this regime.  Modulo
this caveat, \muf\ provides a fairly viable platform to study the
radial distribution of galaxy properties around central galaxies.

\section {Conformity in {\tiny s}SFR, \HI\ richness and colour of the galaxies.} \label{gal_conf}

Traditionally, galaxy conformity is known as the tendency for central
galaxies and their satellites to have similar
colours~\citep{Weinmann-06-a}.  However, conformity could in principle
be associated with any galaxy property.  For instance, conformity
has recently been quantified in neutral hydrogen~\citep{Kauffmann-13},
with central and satellite galaxies found to be similar in their
\ion{H}{i} richness.  One could equivalently define conformity
between different galaxy properties.  For example, one could quantify
by how much blue central galaxies have higher \ion{H}{i} satellites,
or quantify how older central galaxies have lower sSFR satellites.
We will call this {\it cross-conformity,} differentiated from {\it
auto-conformity} (or just, conformity).

In this section, we examine galaxy conformity in \muf\ in terms of
colour, \ion{H}{i} content and sSFR.  We also make predictions for
cross-conformity among these properties.  Here we will take bins
of central galaxy stellar mass in order to be able to compare to
observations particularly of \citet{Kauffmann-13}, but we forego a
detailed matching of selection for particular samples that can be
critical for proper quantitative interpretation (as discussed in
\S\ref{intro}).  Instead, we focus on the nature and strength of
conformity as predicted in \muf.

Using the approach of Kauffmann modulo our definition of central
galaxies, the closest neighbours are satellites, representing
one-halo conformity, and those farther out are other haloes' central
and satellite galaxies representing two-halo conformity.  Hence for
each neighbour property, we generate a plot of three mass bins,
which we show in columns, and three central galaxy properties, which
we show as rows.  We will consider conformity in the neighbour
properties of colour, sSFR and \HI\ richness; thus we have three
such plots.

\begin{figure*}
\includegraphics{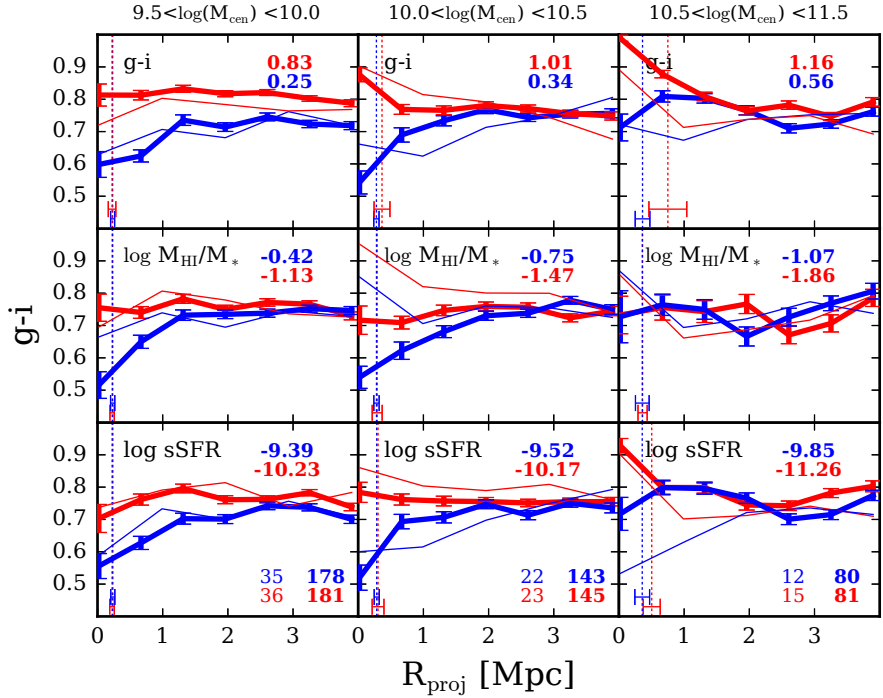}
\caption{
Median {\it g-i} colours of galaxies as a function of projected distances
around central galaxies divided in three stellar mass bins (columns).
Within each stellar mass bins, the red/blue curves denote the median colours
of galaxies around the upper/lower quartiles of central galaxies in
g-i (top panels), HI richness (middle panels), and sSFR (bottom panels).
The median values of each quartiles of the central galaxies
are shown on top of each panel.
The number of central galaxies in each stellar mass bins are shown
on the bottom right of the bottom panels: the lower (upper) quartile
is in red (blue) (normal font: m25n512, bold font: m50n512).
The  thick lines show our fiducial box (m50n512) and the thin lines our smaller
(higher resolution) box (m25n512).
The errorbars are estimated with jackknife resampling.
The vertical dashed lines show the median values of the $R_\mathrm{vir}$
of the hosts of the central galaxies (colour coded),
with the errorbars at the bottom left quantifying the standard deviation.
The top row represents the conformity signal in {\it g-i}, while 
the bottom two rows represent the cross-conformity of centrals' \HI~
richness and sSFR with satellites' {\it g-i}.
}
\label{fig_comp_3}
\end{figure*}

\begin{figure*}
\hspace{-0.5cm}
\includegraphics{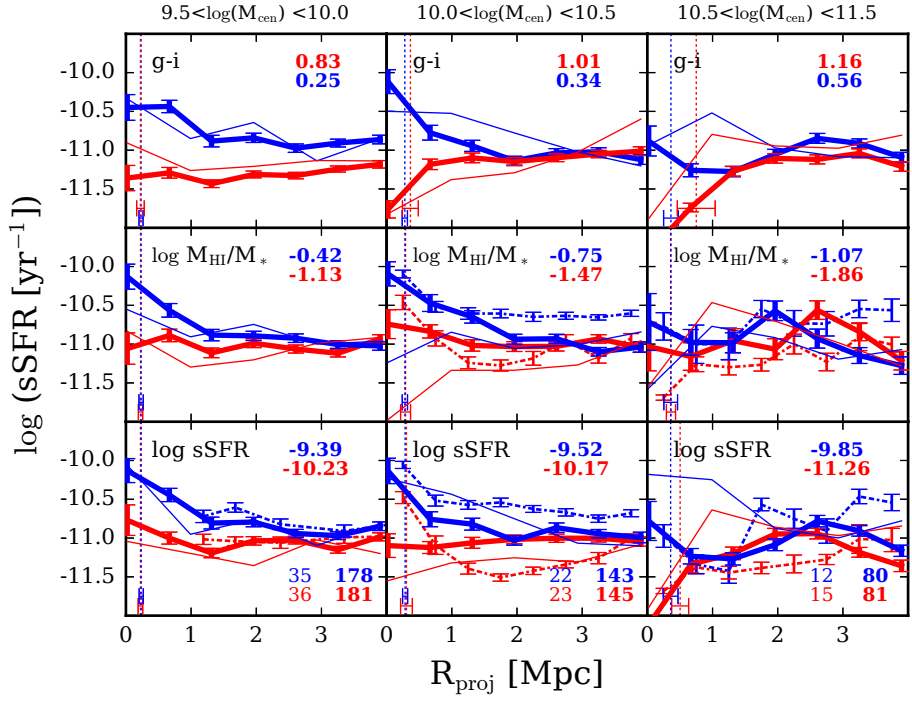}
\caption{Similar to \ref{fig_comp_3}, except showing the sSFR
of the neighbours. Dashed lines (colour coded) show SDSS observations from
\citet{Kauffmann-13} and \citet{Kauffmann-15} for reference,
with the caveat that we have not mimicked their selection in detail.}
\label{fig_comp}
\end{figure*}

\begin{figure*}
\includegraphics{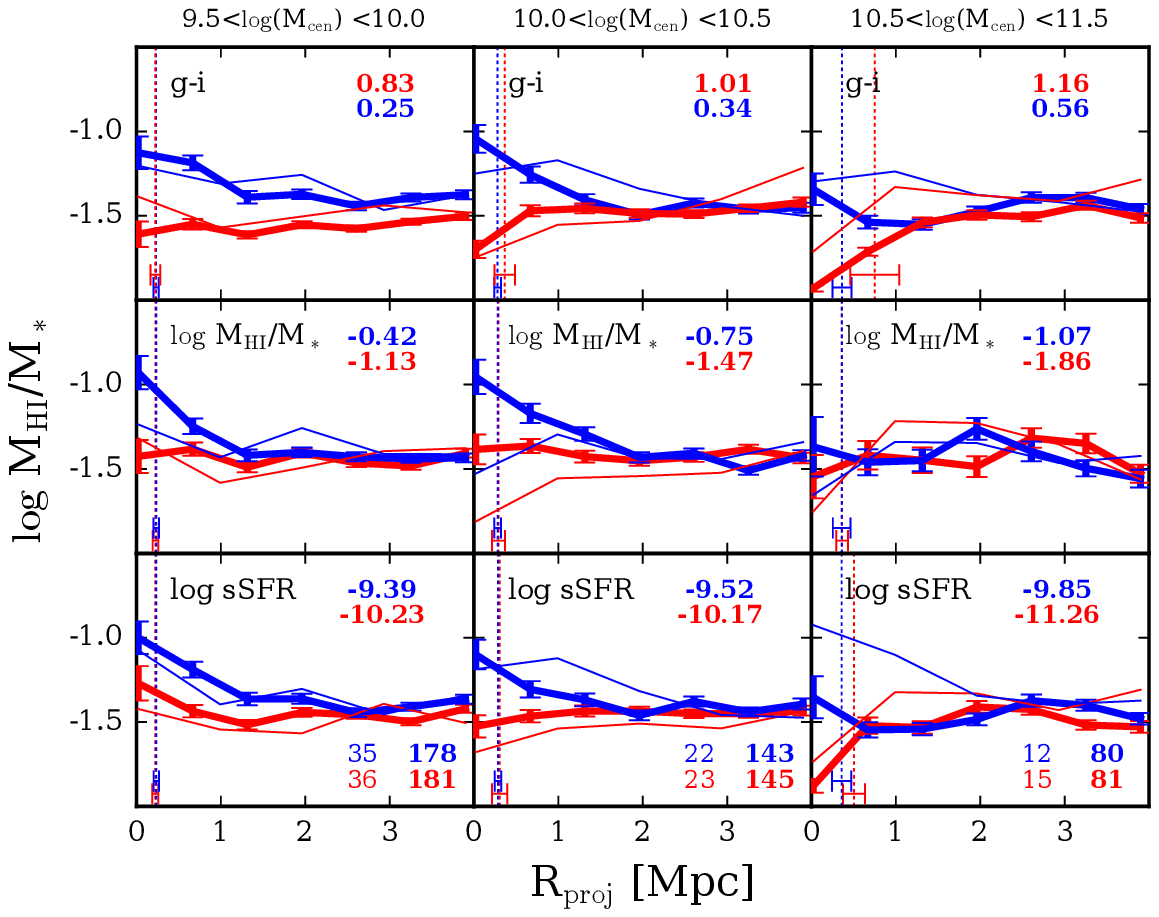}
\caption{Similar to \ref{fig_comp_3}, except showing the \HI~richness
of the neighbours.}
\label{fig_comp_2}
\end{figure*}

Figure \ref{fig_comp_3} shows the first of these plots, depicting
the $g-i$ colour of the neighbouring galaxies as a function of
projected distance to their respective centrals $R_{\rm proj}$.
The error bars are from jackknife resampling. The top row represents
traditional conformity: the tendency for neighbouring galaxies to
share the colour of their central galaxy. The median colour of
central galaxies in the top and bottom quartiles of colour are shown
as the values in the upper right, for each mass bin.  The
values at the bottom right of the bottom panels show the number of
central galaxies for each stellar mass bins (color coded,
where the normal (bold) fonts are for m25n512 (m50n512)).  The
thick lines show our fiducial box (m50n512), but we also show our
smaller (higher resolution) box (m25n512) with the thin lines for
comparison. The vertical dotted lines (colour coded) show the
median values of the virial radii of the central galaxies with
$1\sigma$ uncertainties shown towards the bottom of each panel.
We show the plot out to $R_{\rm proj}=4$~Mpc, beyond which
point we will later show the conformity signal essentially disappears
in all tracers.

In the lowest mass bin, conformity is evident at all scales; the
central galaxies have a median $g-i=0.83$ magnitudes, and the
neighbours tend to have a similar colour all the way out to 4~Mpc.
The colour of the bluest quartile of centrals is $g-i=0.25$, but
here the neighbours are generally redder than the central, with
$g-i\approx 0.6-0.7$; nonetheless they are still clearly bluer than
the neighbours of red centrals.

The intermediate and high-mass bins show conformity as well, but
restricted only to within $R_{\rm proj}\la 1-2$~Mpc, with a trend
for less extended conformity in larger galaxies.  The conformity
is also relatively weak, as the colour difference in the neighbours
is $\la 0.2$ magnitudes, while the colour difference between the
centrals in the highest and lowest quartiles is much larger, $\sim
0.5-0.7$.

Thus \muf\ clearly shows evidence for colour-colour conformity.
While it weakens with projected distance, it still extends well
beyond $R_{\rm vir}$, and thus we predict two-halo conformity to
exist albeit with a strength that rapidly diminishes with radius.
We will quantify the strength of conformity in \S\ref{gal_conf_ev}.
It could be that the two-halo conformity reflects only backsplash
satellites whose orbit takes them beyond the nominal virial radius,
or neighbourhood-quenched galaxies~\citep{Gabor-Dave-15}; we will
investigate the nature of these galaxies in detail in future work.

The middle row shows the level of colour conformity when selecting
central galaxies by \HI\ fraction.  Cross-conformity between \HI\
richness and colour is equivalently evident as for colour-colour
(auto-)conformity. Here, at low masses the cross-conformity only
extends to $\sim 2$~Mpc, which is similar to what is seen at
intermediate masses.

Finally, in the bottom panel, we show the cross-conformity between
sSFR and colour.  For an extinction-free stellar population, these
two would essentially be identical.  We have included extinction
in computing our colours, so the similarity in trends with the
colour-colour conformity is not completely trivial, but nonetheless
the trends do look quite similar.  As such, herein we will primarily
consider sSFR as a proxy for colour.

An interesting note is that for low and intermediate mass galaxies,
the bluest, most \HI-rich, or most star-forming galaxies all have
close neighbours that are significantly bluer than faraway neighbours.
However, in the most massive bin, this turns around, and the satellite
colours become redder moving in from $R_{\rm proj}\la 1$~Mpc.  The
trend in colour is also present but somewhat weaker than in sSFR,
since it is partially offset by decreasing extinction from the lower
gas content at small radii.  This shows the strong effect that our
halo-based quenching model has on truncating the gas content and
star formation in satellite galaxies.

In Figure~\ref{fig_comp}, we show conformity in the sSFR's of
neighbouring galaxies. The trends are quantitatively similar to
that seen in the colour conformity case, except flipped in sign
since higher colours correspond to lower sSFR.  Again, conformity
is seen out to $\sim 2$~Mpc in lower-mass galaxies, and is strongly
radial dependent.  Here, we can see that the sSFR conformity clearly
increases at low separations for \HI\ and sSFR, for low and
intermediate mass galaxies.

Conformity has been measured in SDSS with respect to sSFR
\citep{Kauffmann-13} and \HI~\citep{Kauffmann-15}.  These data are
shown as the dashed lines in the lower two rows.  The median values
of the simulated galaxy properties are in reasonable agreement with
the observed trends.  At low and intermediate central masses, the
amplitude of the separation between the top and bottom quartiles
are in agreement with data at $R_{\rm proj}\la 1$~Mpc, and also
show an increase of neighbours' sSFR to small separations.  For the
largest masses, the upturn at small radii in neighbour colour, \HI\
content, and sSFR is not as strong, and this again is in broad agreement
with observations. We emphasise that our isolation criterion based
on using only central galaxies in friends-of-friends haloes is not the
same as that used in \citep{Kauffmann-13}. 
As we showed in Figure~\ref{mass_ratio}
there are only mild differences between these different isolation criteria
and strongest only for the most massive centrals, and thus it is instructive to 
compare to the \citep{Kauffmann-13} data even if the criteria are
not identical.  Meanwhile, our results provide predictions that are
more comparable to recent group catalog-based analyses such as
that of \citet{Tinker-17}.

In detail, at intermediate masses the observed neighbours of
lowest-quartile centrals become substantially higher (bluer) at
small separations, while \muf\ does not predict this trend (though
the agreement is good for highest-quartile centrals).  One explanation
for this discrepancy could be that mergers increase sSFR at low
separations, but this enhancement is not properly reflected in the
simulations, perhaps owing to resolution.  At larger radii, the
data continues to show strong conformity that is not seen in \muf,
but these data are somewhat in doubt owing to interloper contamination
as discussed in \citet{Sin-17,Tinker-17}.  Our results are broadly
in agreement with the semi-analytic or abundance matching models
presented in those works, for which conformity disappears at distances
not far beyond the virial radius in more massive galaxy samples.

For completeness, we consider in Figure~\ref{fig_comp_2} the
conformity trend in the \HI\ richness of neighbours, for central
galaxies split by colour (top row), \HI\ richness (middle), and
sSFR (bottom).  The trends generally mimic those in the other cases,
in that conformity is only detectable out to $R_{\rm proj}\la 2$~Mpc,
and is only strong at $R_{\rm proj}\la 1$~Mpc.  The neighbours'
\HI~ richness, like colour and sSFR, becomes higher (bluer) towards
small separations for low-mass centrals, but become lower (redder)
for high-mass centrals.

Comparing the $50\hmpc$ (thick) and $25\hmpc$ (thin) line
predictions, we see that in general the conformity signature is
qualitatively similar at both resolutions, though they are not
within each others' formal error bars.  As shown in \citet{Dave-17a},
the two volumes are not completely converged in terms of many of
their stellar and gas properties.  Ironically, the convergence in
conformity strength appears to be best-converged for the smallest
centrals; at higher masses, the trends are less well mimicked in
the two volumes, though overall the trends are similar.  
On top of the rather small sample from m25n512,
we also speculate that the discrepancy might also be from
the model behaving differently at different resolutions.
We note that the error bars for the $25\hmpc$ results (not shown for clarity)
are generally larger than for the $50\hmpc$.

Overall, \muf\ displays fairly strong galaxy conformity within $\la
1$~Mpc projected radius, and weak conformity in most cases out to
$\sim 2$~Mpc, which disappears at high masses.  The conformity
signature is present and similar in all permutations of central vs.
neighbour galaxy properties considered here, namely colour, \HI\
richness, and sSFR.  The trend to small radius at large central
masses shows the impact of halo quenching, but for lower central
masses we see bluer, more gas rich, and higher sSFR neighbours towards
small radii.  Comparing to observations of
\citet{Kauffmann-13,Kauffmann-15}, the trends generally agree at
projected radii less than about 1~Mpc.  \muf\ does not produce as
strong conformity at larger scales as inferred by those data, but
at large separations these data may be impacted by interloper
contamination.  Observational comparisons are sensitive to details
of selection effects etc. particularly at large radii, so we consider
the outcome of conformity in \muf\ and the broad agreement at $R_{\rm
proj}\la 1$~Mpc to be encouraging, and leave a more detailed data
comparison for future work.


\section{The Nature of Conformity} \label{gal_conf_ev}

In the previous section we showed that galactic conformity is present
in \muf\ in specific star formation rate, \HI\ richness, and colour
of galaxies at similar levels, with conformity being stronger at
small separations and extending farther out in lower-mass central
galaxies.  Physically, one might regard conformity as a reflection
of quenching processes associated with hot massive
haloes~\citep[e.g.][]{Peng-12,Gabor-Dave-15}.  \muf\ assumes a
quenching halo mass scale of $\sim 10^{12}M_\odot$, and hence the
effects of conformity might be expected to be stronger in haloes
above this mass scale, since the gas surrounding satellites has now
been forcibly heated to the virial temperature.  However, starvation
and stripping processes can happen in lower-mass haloes through
tidal interactions and harrassment.  Hence it is an interesting
question to quantify how conformity changes with halo mass,
particularly across our quenching mass threshold.

To this end, in this section we subdivide our sample with respect
to halo mass above and below our nominal quenching scale, as opposed
to subdividing in stellar mass as in the previous section in order
to more closely compare with data.  Furthermore, we consider
conformity in a wider range of properties beyond only what has been
observed, to a more exhaustive set of galaxy properties.  This will
help us identify which properties display the strongest conformity,
and hence in some sense drive galactic conformity.

\subsection{Conformity in non-quenched haloes}

\begin{figure*}
\includegraphics{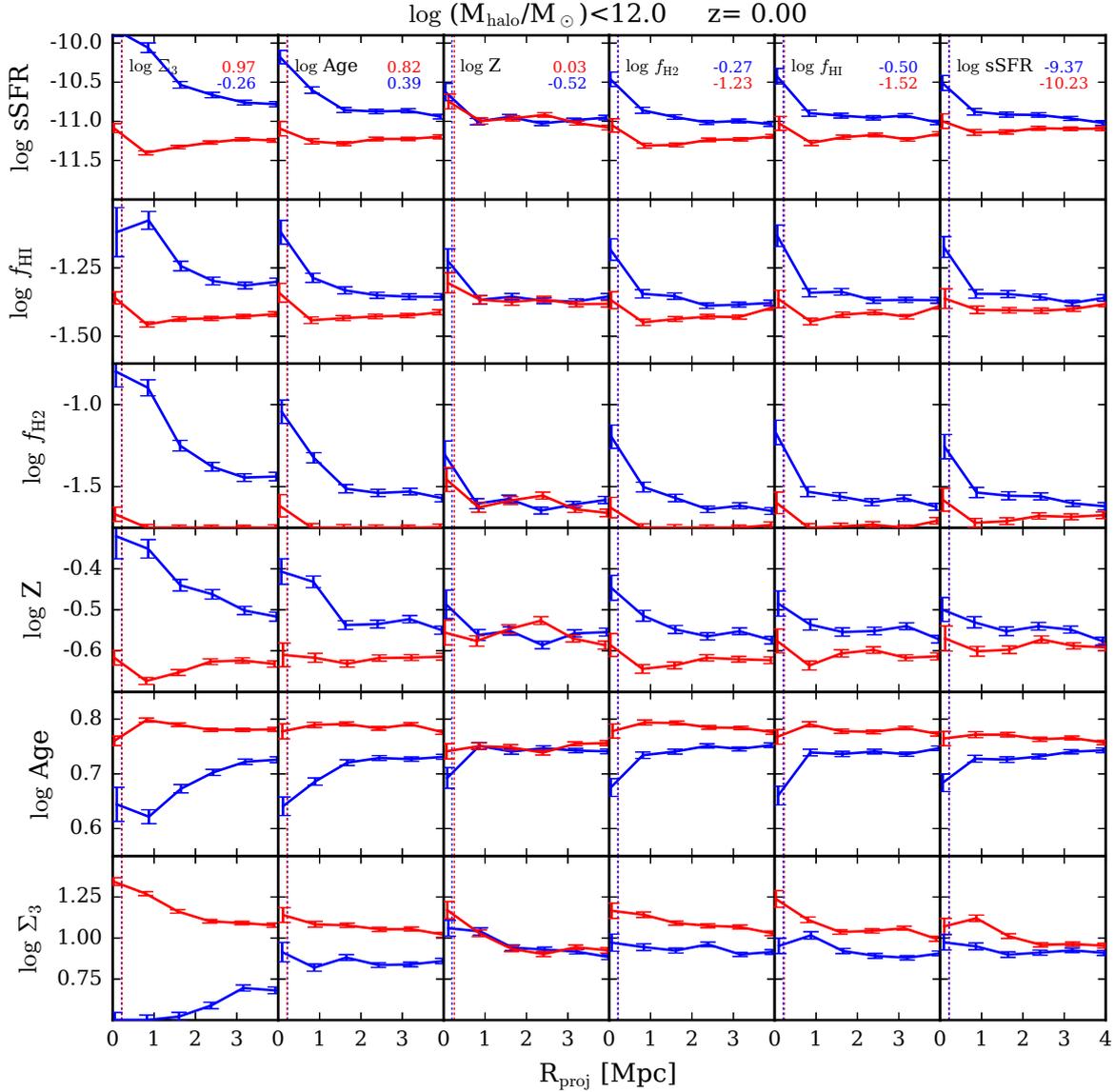}
\caption{Neighbour galaxy median properties versus projected distance
to their central galaxies living in haloes with mass
M$_\mathrm{halo,cen}<10^{12}M_\odot$,
binned by central galaxies in highest and lowest quartile
for each given property. The specific property shown in each column is
indicated in the upper left of the top row, while the median quantities
for the central galaxies are indicated on the upper right
(color coded with the lines).
The dashed vertical lines are the median values of the $R_\mathrm{vir}$
of the central galaxies in the highest and lowest quartiles.}
\label{env_small_halo}
\end{figure*}

Figure \ref{env_small_halo} shows, from top to bottom panels, the
sSFR (yr$^{-1}$), $f_\mathrm{HI}$ (\HI~richness), $f_\mathrm{H_2}$
(molecular hydrogen fraction),  $Z$ (gas phase metallicity), Age
(median value of the stellar ages, in Gyr) and $\Sigma_3$ (third
projected nearest neighbour density) of central galaxies in haloes
of $\log (M_{\mathrm{halo}}/\msolar)<12$, ordered by each of those
properties (columns). The trends for sSFR and $f_\mathrm{HI}$ were
shown in Figure \ref{fig_comp} and Figure \ref{fig_comp_2} binned
by central stellar mass, but here we show them binned by halo mass.
The diagonal panels (from lower left to upper right) show the
(auto-)conformity in each quantity, while the off-diagonals show
the cross-conformity with the rows representing the neighbour
properties and the column representing the central properties.

Before we delve into a detailed analysis, it's important to keep
in mind the nature of galaxies in this halo mass bin.  Note that
the median sSFR of the reddest quartile of central galaxies is
$\log$ (sSFR/Gyr$^{-1}$)=$-1.23$, which is not a quenched galaxy
by our (or almost any reasonable) definition.  This is not surprising,
since in \muf, we do not apply our quenching prescription in this
range of halo masses.  Nonetheless, it is worth bearing in mind
that here we are mostly examining trends among star-forming galaxies,
subdivided into redder vs. bluer, more gas-rich versus less, etc.
Since there exist trends with galaxy mass in these
quantities~\citep[e.g.][]{Dave-17a}, binning in these quantities
implicitly includes some trend with galaxy mass.  Ideally, one would
remove this effect by sub-binning in narrow bins of galaxy (or halo)
mass, but given our statistics, this is not feasible.

Let us consider the auto-conformities first, along the diagonal
panels.  For all quantities except $Z$, the conformity is well-defined
out to remarkably large scales, exceeding the 4~Mpc (projected)
limit that we consider here, with a strength that diminishes
relatively slowly.  We also saw this behavior in some but not all
cases around the low-mass centrals shown in
Figures~\ref{fig_comp_3}-\ref{fig_comp_2}.  This implies that
neighbours know about their large-scale environment out to quite
large distances.  This likely does not reflect halo (AGN) quenching,
ram pressure stripping, or other processes traditionally associated
with massive haloes, since it is not confined near the haloes
themselves, nor are the selected massive haloes.  Instead, it
represents conformity driven by the growth of large-scale structure
by the tendency for massive galaxies (with e.g. redder colours and
lower gas content) to live around other massive galaxies.

The most striking trend is seen in the first column of Figure
\ref{env_small_halo}.  Here, we see that the difference between the
satellite properties is always greatest when the centrals are
subdivided by our environment measure $\Sigma_3$.  This is true not
only for the $\Sigma_3$ auto-conformity but for every cross-conformity
measure as well.  From this, we conclude that the most important
driver of conformity is the nearby galaxy density, and clearly
identifies galactic conformity in low-mass haloes as environmentally-driven.
This is perhaps unsurprising, since conformity is inherently itself
a tracer, i.e. if galaxy properties vary smoothly with environment,
then having many galaxies of a similar type close to each other
will give rise to strong signals in both $\Sigma_3$ and galaxy
conformity.

The next strongest level of difference between satellite quartiles
is provided by separating centrals in stellar age (second column
from left).  This is qualitatively consistent with the findings of
\citet{Hearin-16}, who argued that assembly bias is crucial for
driving two-halo conformity.  Although we have restricted this
sample to $M_{\mathrm{halo}}<10^{12} \msolar$, it is possible that
the environmental dependence of the halo mass function within this
mass range may drive conformity, with the age aspect being a
consequence rather than a driver~\citep{Zu-Mandelbaum-17}; this
would be consistent with our results that the environment shows the
strongest trend in conformity.  To test this, we would need to
construct an age-matched sample within each environment, but
unfortunately given our limited simulation volume, the results are
too noisy to extract clear trends.

We next examine the gas and star formation rate conformity properties.
Interestingly, the conformity signature is at least as strong when
centrals are subdivided by $f_{H2}$, as compared to either sSFR or
$f_{HI}$.  This is notable since current observations have mainly
probed the latter two (since they are the most observationally
accessible given current data), but \muf\ predicts that these
conformity signals are actually similar if not weaker compared to
that seen in molecular gas fractions.

Finally, the metallicity has a very minor conformity signal, only
at R$_\mathrm{proj} <~ 1$~Mpc.  The most curious aspect is that
qualitative trend in the metallicity of neighbours relative to the
sSFR.  Looking at the top row of the metallicity column, then at
low R$_\mathrm{proj}$, one sees that low-$Z$ centrals tend to have
neighbours with slightly higher sSFR.  This is consistent with the
fundamental metallicity relation~\citep[e.g.][]{Mannucci-10,Lara-Lopez-10},
where at a given mass, galaxies with high sSFR have low-$Z$ and
vice versa; this trend is reproduced in \muf~\citep{Dave-17a}, and
apparently extends to satellites as well.  In contrast, if you look
at the sSFR (6th) column, fourth panel down, it curiously shows
that centrals with high sSFR are surrounded by neighbours with {\it
high} metallicity -- in other words, the trend is flipped, and
metallicity actually shows anti-conformity!  The differences are
subtle but robust, which we will quantify later on.  This may arise
as an age effect -- one can see in the Age (2nd) column, fourth
panel down, that galaxies with young ages tend to be much more
metal-rich, because they have had longer time to form stars and
hence enrich themselves.

In general, we note that the overall amplitude of conformity looks
most similar in the columns of our plot, rather than in the rows.
This means that the conformity is generally most driven by the {\it
central} galaxy property being examined, but relatively independent
of the neighbour property being examined.  In contrast, the trends
with radius are most similar when examining a particular neighbour
galaxy property.

The physical interpretation of one-halo conformity in non-quenched
haloes likely traces back to halo assembly bias.  This is because,
in this regime, there is no explicit physics that will turn nearby
galaxies red, as there is in the quenched haloes case where there
is strong local quenching feedback.  Hence the similarity of galaxy
colours likely arises owing to the tendency for haloes living in
denser regions to have formed earlier and be surrounded by more
gravitationally shock-heated gas, which results in an overall
reduction of the accretion rates onto all galaxies living in such
extended structures.  This is essentially a form of assembly bias,
in which galaxies at a given halo mass experience different growth
histories depending on their large-scale assembly history.  Such
an interpretation is consistent with that of \citet{Tinker-17} from
SDSS and \citet{Bray-16} from examining the origin of conformity
in the Illustris simulation.

In summary, $M_{\mathrm{halo}}<10^{12}\msolar$ haloes show noticeable
conformity across almost all galaxy properties, with the strongest
absolute differentiations among neighbours occurring in environment
($\Sigma_3$).  Specific SFR, $f_\mathrm{HI}$ and Age have actually
rather weak conformity signal relative to that, and somehow comparable
to molecular fraction $f_\mathrm{H2}$. There is little conformity
in galaxy metallicities, and in fact shows opposite trends depending
on whether one considers cross-conformity versus the central's
metallicity or the neighbour's metallicity.  Some of these trends
may arise owing to mass trends among central galaxies when binned
into upper and lower quartiles.  We will discuss conformity more
quantitatively in \S\ref{gal_conf_ev}.

\subsection{Conformity in quenched haloes}

\begin{figure*}
\includegraphics{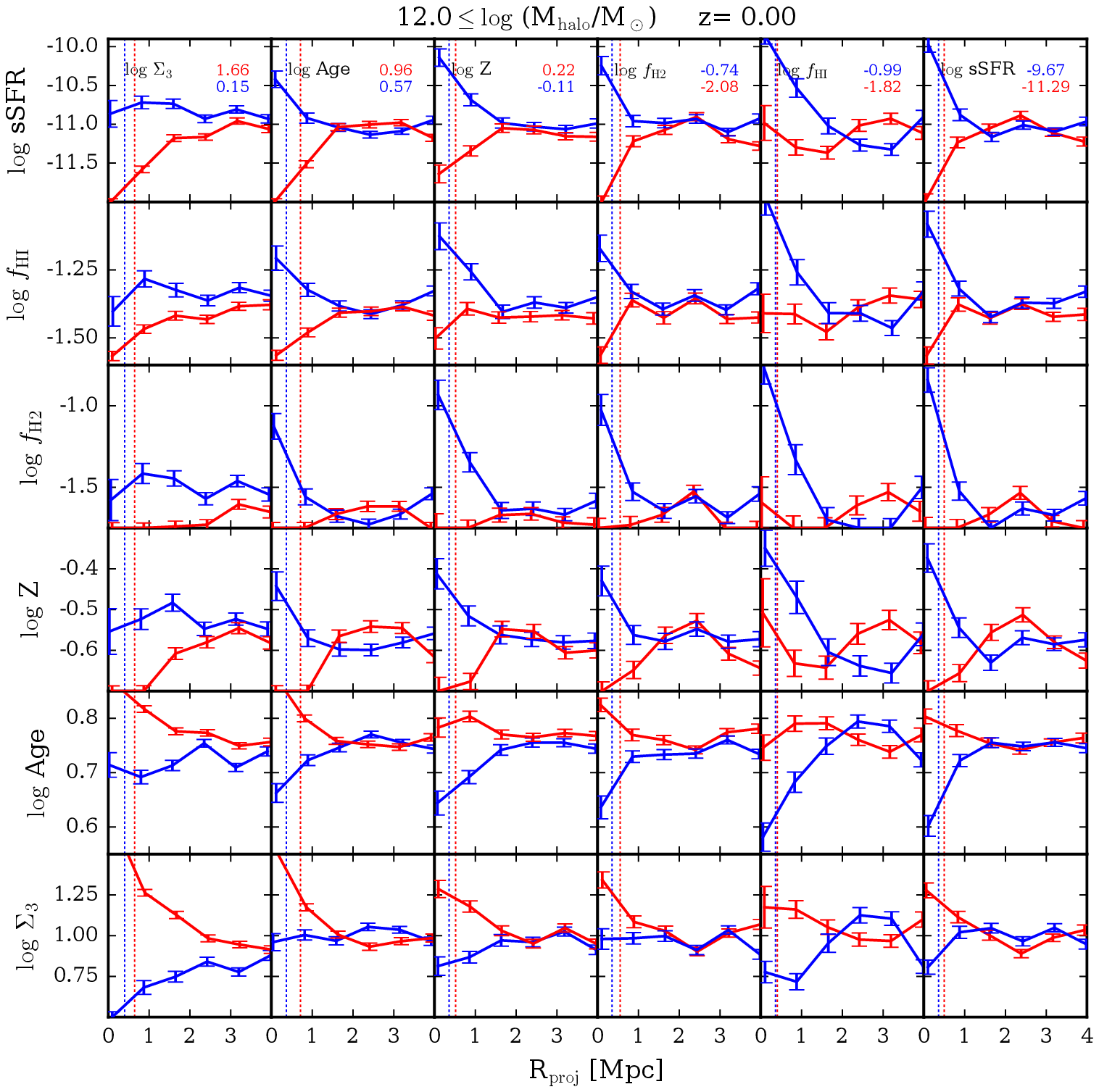}
\caption{Similar to Figure \ref{env_small_halo} except for centrals living in
M$_\mathrm{halo,cen}\geq10^{12}\msolar$.}
\label{env_massive_halo}
\end{figure*}

We now conduct a similar investigation, but for centrals within
quenched haloes having $M_{\mathrm{halo}}\geq10^{12}M_\odot$.
Figure~\ref{env_massive_halo} shows the analogous plot to
Figure~\ref{env_small_halo} for these quenched haloes.  Note that
the typical sSFR of upper-quartile centrals is only a factor of 2
lower than in the low-mass halo case, but for lower-quartile centrals
the sSFR is more than an order of magnitude lower, showing the onset
of a considerable quenched population in massive haloes. Similarly,
the magnitude of the difference is substantially larger between the
highest and lowest quartile in gas content as well.

The trends at high halo masses are clearly different than for the
lower mass haloes examined in the previous section.  Most strikingly,
conformity in just about every property is now restricted only to
R$_\mathrm{proj} \la \sim1-1.5$~Mpc, with the exception of $\Sigma_3$.
Within that range, the radial trend is much stronger than in the
low-mass halo case, as quenching clearly plays a role in impacting
the satellite galaxy population.  The one-halo conformity is obviously
well stronger than the two-halo conformity, showing that satellites
in particular are very strongly impacted by environmental-specific
processes owing to quenching.

Looking at individual properties, we see once again that environment
($\Sigma_3$) shows the strongest absolute levels of conformity, and
its auto-conformity extends out to $\sim 4$~Mpc.  The cross-conformities
tend to nearly vanish beyond $\ga 2$~Mpc.  But in contrast to the
low-mass halo case, the strengths and radial trends of the conformities
in other properties are remarkably similar, and even the metallicity
shows strong conformity.

These trends show qualitatively the behaviour one would expect, with
centrals that are high-sSFR, gas-rich, young, and low-density having
like neighbours.  The exception again is metallicity, where low-$Z$
centrals tend to have high-$Z$ satellites and neighbours, and vice
versa.  The anti-conformity of metallicity is an interesting testable
prediction.

Recall that in \S\ref{sat_mf} we highlighted an issue with these
simulations in that they overproduce the number of low-mass quenched
satellites.  This is expected to increase the 1-halo conformity
term for the reddest (and analogously least star-forming and gas-rich)
quartile, since quenched central galaxies are also red.  The magnitude
of this effect is difficult to quantify without a new feedback model
that is able to mitigate this discrepancy.  We are currently working
on such a model, but do not have final results at this time.  For
now, we note that the quenched halo predictions may change depending
on the new input physics required to fix this discrepancy.

In summary, both low-mass and high-mass (quenched) haloes show
conformity in virtually all quantities, but the trends are qualitatively
different.  High-mass haloes show a relatively confined (spatially)
extent of conformity, with a very strong radial trend, and similar
conformity strength in all properties except for $\Sigma_3$ which
is the strongest.  In the next section we discuss our approach to
quantifying these trends, in order to more carefully inter-compare
and study their evolution with redshift.

\begin{figure}
\hspace{-0.5cm}
\includegraphics{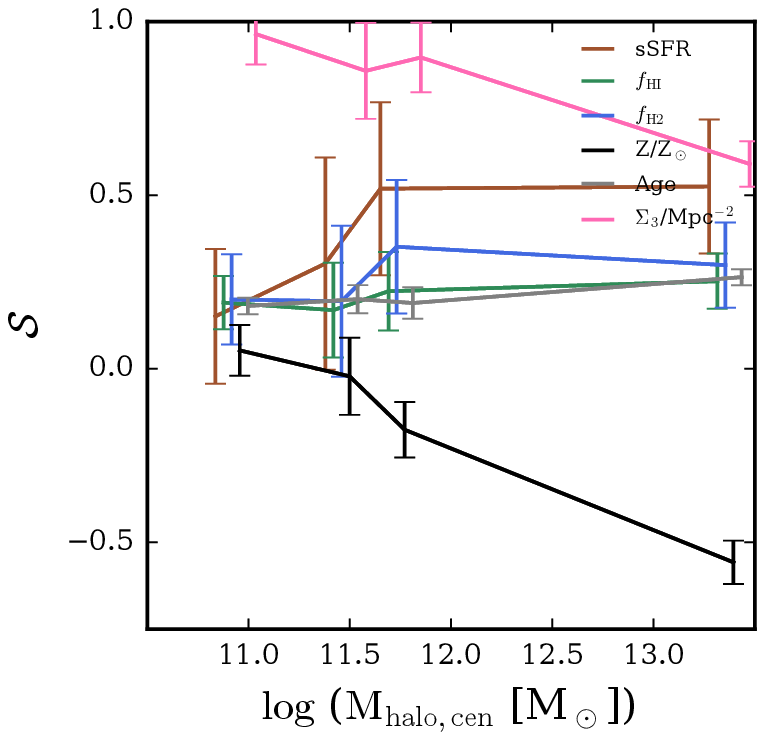}
\caption{Halo mass dependence of the auto-conformity strength in
\muf\ at redshift $z=0$.  Halos have been binned so that each bin contains
approximately the same number.  $1\sigma$ error bars are derived
by jackknife resampling.  Environment shows the strongest conformity
at low masses, but gets weaker with halo mass.  sSFR shows the
strongest conformity among the remaining quantities, and increases
with halo mass from $M_{\rm halo}\sim 10^{12}M_\odot$.
Metallicity displays anti-conformity.}
\label{fig_rad_1}
\end{figure}

\subsection{Quantifying conformity}

Previous works have generally focused on quantifying conformity by
measuring its detectable radial extent, within a given tracer.  The
next logical step would be to quantify the conformity strength in
a manner such that we can intercompare the strengths amoung various
tracers.  In the previous sections, we have focused on comparing
the {\it absolute} strength of conformity within a given neighbour
property, binned by central property.  However, it is not obvious
how to compare this strength between different properties, since
effectively this absolute conformity signal (say, the difference
between the red and blue curves) has units associated with that
property.  Thus we need a new a measure that also enables
cross-comparisons between various neighbour properties, in order
to more robustly determine which neighbour property shows the
strongest conformity.

The fundamental idea of conformity is to quantify how well the
neighbours follow the trends of their central.  In this sense, a
good measure of conformity to inter-compare properties would be to
measure the difference between the neighbour properties, relative
to how much difference there is in the central galaxy properties.
This may be regarded as a {\it relative} conformity, as opposed to
the absolute conformity that we have investigated in the previous
sections.  As such, we define (relative) conformity strength
$\mathcal{S}$ as follows:
\begin{equation}
\label{strength}
\mathcal{S} =
{1\over N_\mathrm{bin}} \sum_{i=0}^{N_\mathrm{bin}-1}
{Q1_i - Q4_i \over {Q1_\mathrm{cen} - Q4_\mathrm{cen}}}
\end{equation}
where $Q1_\mathrm{cen}$ ($Q4_\mathrm{cen}$) is the property
of the central galaxies in the 1$^{st}$ (4$^{th}$) quartile,
$Q1_i$ ($Q4_i$) is the (logarithm of the) property of the galaxies neighbouring
the 1$^{st}$ (4$^{th}$) quartile in the radial bin $i$.  In this
way, we normalize the difference in neighbour properties by
the difference in the central property.  Generally, we will
consider ${\cal S}(R< 2\;{\rm Mpc})$, i.e. summing all bins 
that are within 2~Mpc of the central galaxy.  We will also
consider ${\cal S}(R)$, i.e. taking a single bin at each radius.

Broadly, $\mathcal{S}=1$ means that the neighbours show exactly the
same difference between the top and bottom quartiles as the centrals;
this is full conformity.  $\mathcal{S}=0$ means no conformity.  We
will characterise weak conformity as $\mathcal{S}<0.5$, and strong above
this.  It is possible to get $\mathcal{S}>1$ which indicates very
strong conformity, with neighbours showing a greater difference in
a property than even their centrals, or even $\mathcal{S}<0$ which
is anti-conformity.  Hence this quantity provides an intuitive
measurement of conformity that is independent of the given property.
We note that this is note a new definition of conformity.
We use similar definition of conformity from \citet{Kauffmann-13} but only
extend it to be presentable in one number.

\subsection{Conformity as a function of mass and radius}

\begin{figure}
\hspace{-0.5cm}
\includegraphics{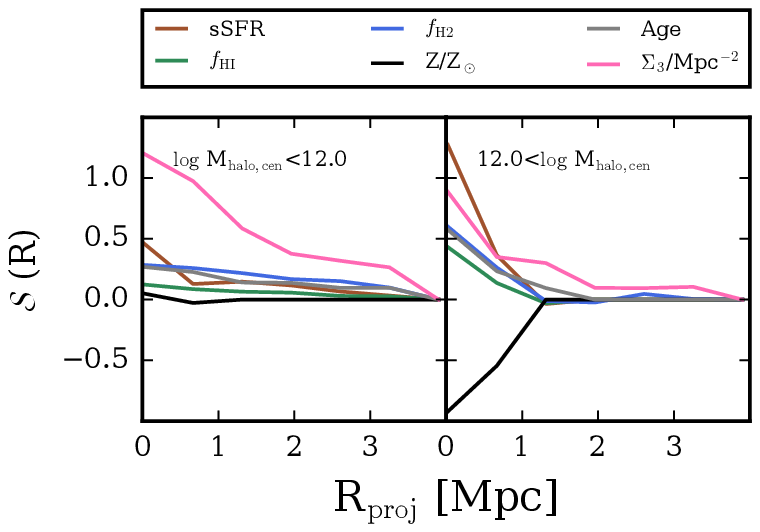}
\caption{Radial dependence of conformity strength ${\cal S}(R)$.  Here
$\cal S$ is computed within each radial bin.  Low-mass haloes show weak
conformity with little radial trend, except for $\Sigma_3$.  Massive
haloes show stronger conformity for neighbours close to the central, but
little conformity at $R_{\rm proj}>1$~Mpc.}
\label{fig_rad_4}
\end{figure}

In Figure \ref{fig_rad_1}, we shows $\mathcal{S}(R< 2\;{\rm Mpc})$
at $z=0$ for central galaxies binned by their halo masses: we chose
each bins to contain the same number of central galaxies. $\Sigma_3$
signal shows the strongest conformity signal, quantifying the trend
in the previous sections that conformity is most strongly related
to environment. The $\Sigma_3$ conformity is however anti-correlated
with halo mass, so environment is the primary driver at low and
intermediate halo masses, but at the highest halo masses it is
comparable to the other quantities.  While we discussed previously
that this perhaps not surprising since conformity corresponds to
the similarity in properties of nearby galaxies, while $\Sigma_3$
is itself a measure of how many nearby galaxies there are,  it is
not a trivial result, since it indicates that galaxies that lie
particularly close to each other in dense regions have similar
properties.  Moreover, the trend with mass is interesting, and is
driven by the difference between the central galaxy properties,
i.e. the denominator of $\mathcal{S}$, increasing.  Hence, effectively,
large-scale structure is the primary driver of conformity when
quenching processes are not present, but once they are, then it is
no longer so dominant.

Besides $\Sigma_3$, sSFR clearly shows the strongest conformity at
most halo masses.  Recall that in the previous section we found
that Age and $f_\mathrm{H2}$ were stronger in an absolute sense.
However, when computing $\mathcal{S}$, we normalize this to the
difference in the central galaxy property, and in this case we
discover that sSFR conformity is stronger than that of Age,
$f_\mathrm{H2}$, or $f_\mathrm{HI}$. Thus it appears that, beyond
the obvious dependence of conformity on environment, sSFR shows the
strongest levels of similarity between the neighbours and the central
galaxies.  Though we do not show it here, we expect galaxy colour
would show a very similar trend to sSFR, as we found in \S\ref{gal_conf}.

As discussed previously, metallicity shows anti-conformity at most
halo masses, increasingly so to larger halo masses.  Recall this
is the SFR-weighted gas-phase metallicity, not the stellar metallicity;
one might expect a different trend for stellar metallicity, as
central galaxies within large haloes tend to have metal-rich stars
but their star-forming gas may be dominated by recent infall that
is relatively metal-poor.

Observational analyses from, e.g., \citet{Tinker-17, Sin-17} suggest
that galactic conformity is only prevalent in the relatively nearby
environment, within $\la 1$~Mpc.  It can be seen in
Figure~\ref{env_massive_halo} that, particularly in massive haloes,
that were generally the targets of these studies, the conformity
signal predicted in \muf\ drops quickly with radius.  However, this
is less true in low-mass (non-quenched) haloes
(Figure~\ref{env_small_halo}), which shows overall weaker conformity
but much less radial dependence.

To quantify this, we show in Figure~\ref{fig_rad_4} the conformity
signal as a function of projected radius $R_{\rm proj}$, for
non-quenched (left panel) and quenched (right) haloes.  The qualitative
trends evident in Figures~\ref{env_small_halo} and \ref{env_massive_halo}
are evident here.  The conformity is strongest in the environmental
measure $\Sigma_3$ at all radii, except close in for massive haloes.
Conformity is relatively weak with modest radial dependence in the
low-mass haloes, while it is strong at $R<1$~Mpc in the massive
haloes but mostly nonexistent beyond this (except in $\Sigma_3$).
In fact, in the innermost bin, the sSFR shows stronger conformity
that even $\Sigma_3$, exceeding unity.  It is thus perhaps not
surprising that conformity was first noticed for colours of satellites
around sizeable galaxies~\citep{Weinmann-06-a} -- \muf\ predicts the
conformity signal is quite strong under these conditions.

In summary, conformity is strongest in environment, confirming the
environmental nature of galaxy conformity.  Beyond this, the next
strongest conformity is in sSFR, which is mildly stronger than gas
fraction and Age.  Metallicity presents an odd anti-conformity
particularly in quenched haloes, which may be a signature of recent
accretion of metal-poor star-forming gas into massive central
galaxies.  Massive haloes show a strong conformity signal in most
quantities only at $\la 1$~Mpc, while less massive haloes show a
weaker conformity with a weaker gradient.

\subsection{One-halo vs. two-halo conformity} 

\begin{figure}
\hspace{-0.3cm}
\includegraphics{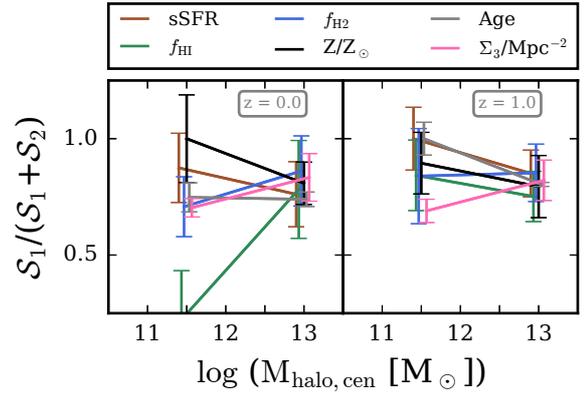}
\caption{Halo mass evolution of the relative contribution of
the one-halo conformity. $\mathcal{S}_1$ ($\mathcal{S}_2$) is
one(two)-halo conformity.  One-halo conformity generally dominates
the signal for all quantities except \HI\ in low-mass haloes.}
\label{fig_rad_2}
\end{figure}

A particularly interesting quantity to examine is the relative
contribution of one-halo versus two-halo conformity to the total
conformity signal, as there has been substantial debate in the
literature regarding the strength, origin, and even existence of
two-halo conformity.  Here we explicitly examine this for each halo,
computing only the conformity associated with galaxies that are
satellites within the haloes of the chosen central, versus those
outside the halo up to 2~Mpc.  Note that this is not a strict radial
cut, since FOF haloes typically have non-spherical shapes.

Figure \ref{fig_rad_2} shows the fractional contribution of one-halo
conformity ($\mathcal{S}_1$) versus two-halo ($\mathcal{S}_2$),
specifically $\mathcal{S}_1$/(($\mathcal{S}_1$+$\mathcal{S}_2$),
as a function of halo mass.  Owing to small number statistics for
the one-halo term particularly at low halo masses, we separate our
central galaxy sample into only two bins of halo mass,
M$_\mathrm{halo}<10^{12}\msolar$ and M$_\mathrm{halo}\geq10^{12}\msolar$.
The left panel shows the $z=0$ results, while the right shows $z=1$.

It is clear that one-halo conformity dominates the overall strength
for almost all quantities for both low and high mass haloes, at
both $z=0$ and $z=1$.  The only deviant case is the \HI~fraction
in low mass haloes, which likely owes to the physical effect that
satellite galaxies around star-forming centrals can have their
\HI~stripped relatively easily, so that such centrals actually end
up having different \HI~fraction relative to their satellites but
more similar to distant galaxies.  Typically, one-halo conformity is
$\sim 3-6\times$ stronger than two-halo conformity, according to our
$\cal S$ measure.  We note that while the strength of
one-halo conformity dominates in most circumstances,
the strength of two-halo conformity at small
radii depends on halo mass, and this may be partly the cause of
divergent results in the literature regarding two-halo conformity.

\subsection{Evolution of conformity} \label{evolution}

\begin{figure*}
\includegraphics{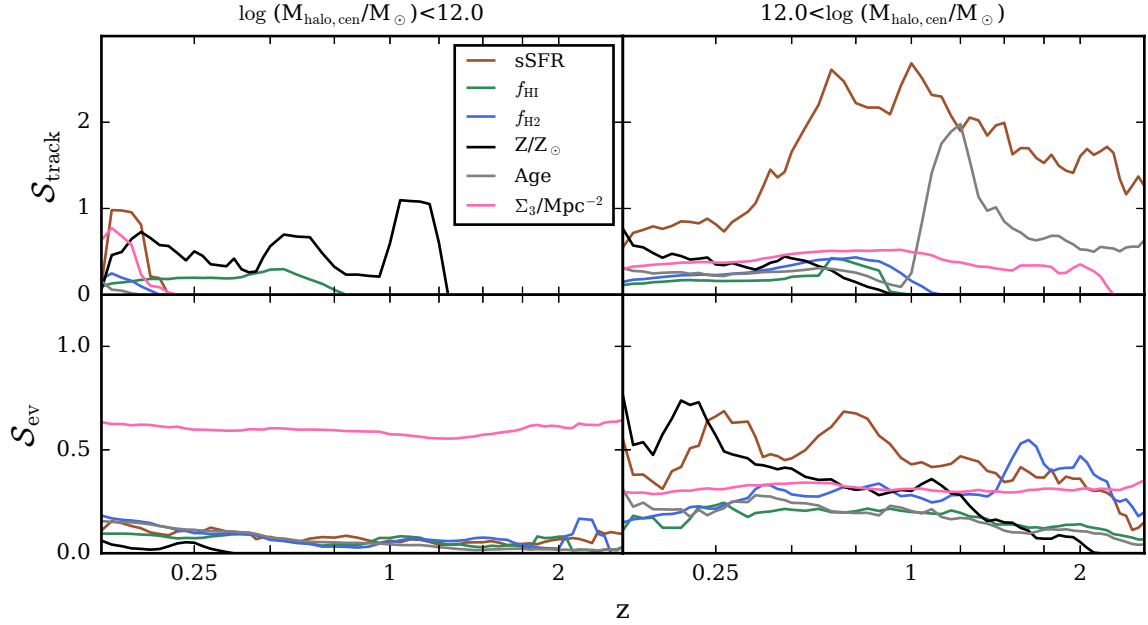}
\caption{Evolution of the strength of galactic conformity. The halo mass
range of the central galaxies are given on top of the columns.
Upper panels show the evolution the conformity strengths tracking back the $z=0$
progenitors in time. Lower panels show the evolution of conformithy within
the listed halo mass bins at each redshift independently.  Conformity is
a late-time emergent phenomenon more associated with high-mass haloes, 
suggesting that it is dependent on quenching physics.}
\label{fig_rad_0}
\end{figure*}

Conformity has been observed to exist out to $z\ga 1$ and beyond,
although the strength of the evolution is difficult to quantify at
this time.  In this section we examine the evolution of conformity
strength $\cal S$ predicted by \muf.

There are two ways to track the conformity back in time for a galaxy
population.  One way is to find the conformity strength for the
most massive progenitors of the $z=0$ central galaxies; we will
call this $\mathcal{S}_\mathrm{track}$, since we are tracking
individual galaxies.  Another way is to consider the centrals in
the same halo mass bins as at $z=0$, separated at $M_{\rm
halo}=10^{12}M_\odot$; we will call this $\mathcal{S}_\mathrm{ev}$,
which shows the overall evolution of conformity.  In both cases we
take the absolute value $|{\cal S}(<2\;{\rm Mpc})|$ for plotting
purposes, and note that only the metallicity shows negative values
(anti-conformity).

Figure \ref{fig_rad_0} shows the evolution of galactic auto-conformity
strength, (i.e quantifying the diagonal panels of Figures
\ref{env_small_halo} and \ref{env_massive_halo}), for central
galaxies binned into two halo mass ranges: $\log
(M_\mathrm{halo}/M_\odot)<12$ and $12<\log (M_\mathrm{halo}/M_\odot)$.
For clarity of the plot, we set the strength to 0 when $\mathcal{S}$
crosses 0 and stay on the other side of the 0-line for two successive
snapshots.  In addition, we smoothed the curves to their moving
averages over $\Delta z=0.1$ for $z<0.5$, and 0.2 elsewhere, to average
over the fluctuations among individual redshift snapshots.

The left panels show $\mathcal{S}_\mathrm{track}$ (top) and
$\mathcal{S}_\mathrm{ev}$ (bottom) for low mass (unquenched) haloes.
For these, conformities are generally weak at all redshifts.  Tracking
our specific central galaxies back from $z=0$, $f_\mathrm{HI}$ and
$Z$ emerge earlier than the others at $z\sim1$, while the others
only appear at $z<0.25$.  Because these are small haloes, we are
limited in how far back we can track these galaxies before they
lack resolution to follow.  However, it is evident that we can track
sufficient numbers till at least $z\sim 1$, so it is interesting
that the low-mass conformity we see at $z=0$ in many quantities is
actually quite a late-time phenomena for these particular galaxies.
This suggests that conformity in e.g. $\Sigma_3$ or sSFR requires
halo masses that approach $\sim 10^{12}M_\odot$,
and at significantly smaller halo masses there is no conformity except
the metallicity anti-conformity.
With larger statistical samples from upcoming larger-volume simulations,
we will be able to test this idea more finely.

Examining $\mathcal{S}_\mathrm{ev}$ for low-mass haloes, we see that
conformity is very uniform and weak at all redshifts, with the
exception of $\Sigma_3$ as noted before.  Again, this suggests that there
is mass threshold for the emergence of conformity that is close to
$\sim 10^{12}M_\odot$.  There is a hint that conformity strength
increases with time for a fixed halo mass sample, but given the
small statistics at $z\gtrsim 2$ it is difficult to draw firm conclusions.

Turning to the high mass haloes (right panels), we see much stronger
levels of conformity in some quantities.  In particular, in
$\mathcal{S}_\mathrm{track}$ (top right) sSFR conformity is extremely
strong out to $z\ga 2$.  Hence for massive haloes, even tracking
them back in time shows that classic (colour-based) conformity
emerges quite early on.  We note that this trend is driven by a few
massive haloes that have a large number of satellites, since
conformity in massive haloes is generally restricted to relatively
small radii and hence are dominated in statistics by satellites.
This also causes some odd behaviour such as the
Age-$\mathcal{S}_\mathrm{track}$ which spikes up around $z\sim 1$;
we hesitate to over-interpret this behavior without more statistics.
It is also the case, as at lower masses, that gas conformity is a
relatively late-time phenomenon, appearing only at $z\la 1$.

For $\mathcal{S}_\mathrm{ev}$ in high-mass haloes, again we see that
the conformity strength is fairly constant with redshift.  The
metallicity (anti-)conformity shows the most significant increase
with time.  This is consistent with the idea that conformity in
most quantities is primarily driven by environment (which correlates
strongly with halo mass), except for metallicity where the
anti-conformity is driven by a different effect namely the late
infall of low-metallicity gas into massive galaxies.

\begin{figure*}
\includegraphics{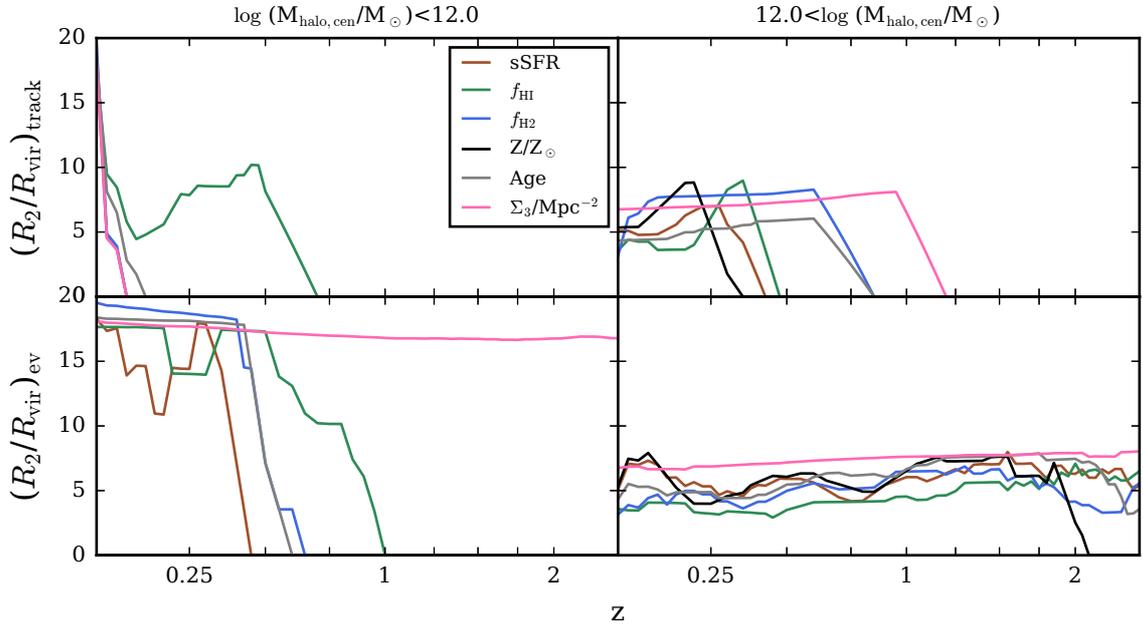}
\caption{Extent of the two-halo conformity. $R_2$ is the distance
from the primary galaxies to the point where the two-halo conformity still exists,
i.e blue and red lines in Figure \ref{env_small_halo} are within each other's
error bars.  At low masses, conformity emerges at late epochs and quickly
to large numbers of virial radii.  At high masses, once it appears, conformity 
is evident out to a fairly constant $\sim 5R_{\rm vir}$.}
\label{fig_rad_3}
\end{figure*}

Another way to quantify conformity is to ask, how far from the
central galaxy should we expect conformity to be evident?  To
quantify this, we measure the distance from the central galaxies
where the two-halo conformity still exists ($R_2$), relative to the
mean $R_\mathrm{vir}$ of the central galaxy subsample.
$R_2$ is measured as the largest projected distance
from the central galaxies where the highest and lowest quartile samples
are first within $1\sigma$ of each other (e.g. where the blue and red lines
are within each other's error bars in Figure \ref{env_small_halo}).
We note that $R_2$ is thus mildly sensitive to the level of the error
bars, and hence the sample size.

In Figure \ref{fig_rad_3} we show $R_2$ as a function of redshift
for low (left panels) and high (right) halo mass bins and tracked
either using progenitors (top panels) or at a fixed halo mass bin
(bottom), analogous to Figure \ref{fig_rad_0}.

At $z=0$ for small haloes, we see as in Figure \ref{env_small_halo}
that the conformity extends out to many virial radii; we truncate
this at 2~Mpc, corresponding to $\approx 18$ median virial radii
in that halo mass bin.
However, tracking these galaxies back in time shows that the conformity
radius disappears extremely rapidly, with the exception of \HI~
fraction for which it extends to $z\sim 2$.  Even using a fixed
halo mass bin, $R_2$ starts dropping beyond $z\ga 0.25$, and disappear
by $z\sim 1$.

For massive haloes, $R_2$ extends typically out to $\sim 5R_{\rm
vir}$, as long as it is present.  Tracking galaxies back, we see
that when conformity emerges, it emerges quickly to this radius.
Of course, this exact radius as well as the exact redshift
depend on the sample size, but the trend of rapid emergence is robust.
When looking within a fixed halo mass bin,
we see the radius to which noticeable conformity
extends is remarkably constant out to $z\sim 2$ in terms of typical
virial radii.

In summary, conformity in low-mass haloes appears to be a late-time
effect, likely arising from haloes as they approach $\sim 10^{12}M_\odot$.
Tracking massive haloes, conformity is evident and strong in sSFR
at all redshifts, but in the gas properties and stellar age it is
relatively weak and emerges at $z\la 1$.  When looking at
$>10^{12}M_\odot$ haloes at each redshift, conformity is fairly
constant in terms of both strength as well as number of radii to
which it extends.  Hence we predict that conformity should be evident
in reasonably massive galaxies at approximately the same level at
higher redshifts as at $z=0$.  These trends are qualitatively
consistent with observations showing that conformity continues to
be evident out to $z\sim 2$.

\section {Summary \& Conclusion}

We have examined galaxy conformity as an emergent property of a
cosmological hydrodynamic simulation of galaxy formation.  In
particular, we employ the \muf\ simulation that has been shown to
reasonably reproduce a range of observed galaxy physical properties
over much of cosmic time~\citep{Dave-16,Dave-17a}.  The approach
for quenching massive galaxies in this simulation is purely heuristic,
utilising a slowly-evolving threshold halo mass above which diffuse
halo gas is prevented from cooling, which well reproduces the
observed red/blue bimodality in the galaxy population~\citep{Dave-17b}
but does not invoke a specific physical model driving the quenching.
Galaxy conformity provides a relatively unexplored statistic with
which to quantify how such quenching impacts the properties of
galaxies in various environments.  To this end, we examine conformity
in galaxy properties that have been previously looked at in the
literature, namely colour, \HI~fraction and specific SFR, along
with a range of other galaxy properties such as molecular content,
stellar age, gas-phase metallicity, and environmental density.

Our main findings are as follows:
\begin{itemize}

\item \muf~yield approximately the observed fraction of total
satellite galaxies as a function of central galaxy mass.  However,
it overproduces the low-mass quenched satellites, most noticeably
around high mass central galaxies when the quenched satellite
dominate the count (Figure~ \ref{fig_sat_mf}).

\item Satellite galaxy colours are generally bimodal for all halo masses
and radii, with quenched satellites at small radii and star-forming
ones farther out.  The transition radius between these varies with
central mass: $<10^{12}M_\odot$ haloes have quenched satellites
dominating only within the inner 10-15\% of $R_{\rm vir}$, while
$>10^{13}M_\odot$ haloes have star-forming satellites primarily in
the outer 10-15\% (Figure~ \ref{fig_color_halo}).

\item Very broadly, galaxy conformity is evident in \muf\ at all
stellar masses examined ($M*>10^{9.25}M_\odot$), in essentially
all quantities examined.  It typically extends significantly beyond
the virial radius, though the strength diminishes with radius
(Figures~ \ref{fig_comp_3},\ref{fig_comp}, \ref{fig_comp_2},).

\item Focusing on conformity in $g-i$ colour, specific star
formation rate (sSFR) and \HI~, we find that conformity at low (central)
galaxy mass is relatively weak, compared to high mass galaxy,
but extends to quite large radii, while conformity at high galaxy
mass declines more quickly with radius.  The cross-conformity among
these three quantities show similar trends to the (auto-)conformity
(Figures~ \ref{fig_comp_3}, \ref{fig_comp}, \ref{fig_comp_2},).

\item We subdivide our galaxies into ``unquenched" ($M_h<10^{12}M_\odot$)
and ``quenched" ($M_h>10^{12}M_\odot$) haloes, which provides a
more direct view of the impact of quenching on neighbouring galaxies.
Low-mass haloes show relatively weak conformity compared
to high-mass haloes, extending to large radius, likely arising
from assembly bias by which galaxies residing in more dense environments
experience earlier growth and more suppression of accretion today.
Meanwhile, high-mass haloes show conformity rapidly declining
with radius and typically disappearing at projected radii above
$\sim 1$~Mpc.  This strong qualitative difference demonstrates that
the presence of a sustained hot halo in our model has a major impact
in driving strong galaxy conformity in our simulation 
(Figures~ \ref{env_small_halo}, \ref{env_massive_halo}).

\item Qualitatively, the strength of the conformity or cross-conformity
signal is most directly correlated with the central galaxy property
being examined, while the radial trend of the (cross-)conformity
signal is dependent on the neighbour property.  Environment (as
measured by the density to the third nearest projected neighbour) and
stellar age appear to have the largest absolute (cross-)conformity
signals, while metallicity shows a very small signal.  The conformity
strength in sSFR, $f_\mathrm{HI}$, and $f_\mathrm{H2}$ are comparable
(Figures~ \ref{env_small_halo}, \ref{env_massive_halo}).

\item We introduce a measure ${\cal S}(R)$ (where $R$ is the projected
radius) to quantify the conformity strength, defined as the difference
in the neighbour galaxies' properties in the first and fourth quartiles
relative to that of the central galaxies.  By normalizing to the
central galaxies, this constructs a dimensionless ``relative"
conformity strength that can be used to inter-compare different
properties.  ${\cal S}(R)$ is unity if the neighbours show as much
difference as centrals, while it is zero if they show no difference,
and can be negative if the satellites' difference is in the opposite
sense to the centrals'.

\item ${\cal S}(R<2 {\rm Mpc})$ is strongest for environment, but
this strength declines with halo mass.  Low-mass ($M_h\sim
10^{11}M_\odot$) haloes show very weak conformity (${\cal S}\sim
0.1-0.2$) in all quantities, but for sSFR and $f_\mathrm{H2}$ this rises
to moderate strengths (${\cal S}\sim 0.3-0.5$) by $M_h\sim
10^{12}M_\odot$.  Age and $f_\mathrm{HI}$ always show relatively weak
conformity.  Metallicity, interestingly, shows increasing anti-conformity
(${\cal S}<0$) with halo mass, possibly owing to the increasing number of
satellites coupled with the reduced gas-phase metallicity in centrals
from small amounts of fresh accretion (Figure~ \ref{fig_rad_1}).

\item Consistent with qualitative impressions, ${\cal S}(R)$ for
low-mass haloes is weak but present at all $R$ in most quantities,
while in massive haloes it is strong at small radii but declines
rapidly and vanishes at $R\ga 1$~Mpc.  At the smallest radius, i.e. for
satellite galaxies, ${\cal S}(R)$ is strongest for sSFR
(or equivalently galaxy colour), and even in low-mass haloes this
declines relatively quickly with $R$ (Figure \ref{fig_rad_4}).

\item For quantities that show significant conformity, ${\cal S}(R)$
is dominated by one-halo conformity over two-halo.  An exception to
this is for $f_\mathrm{HI}$ in small haloes, where small satellites of
gas-rich centrals can get their neutral gas stripped, resulting in
more similarity in this quantity with neighbours farther out
(Figure~ \ref{fig_rad_2}).

\item Tracking conformity back in time for the $z=0$ galaxy population,
we find that conformity is a late-time phenomenon for low-mass haloes
for all properties except metallicity.  Massive haloes develop
conformity earlier, and the sSFR conformity is generally always the
strongest, with ${\cal S}$ significantly exceeding unity at
intermediate redshifts.  These trends are consistent with the idea
that significant conformity only occurs once halo start to be
quenched (Figure~ \ref{fig_rad_0}).

\item Tracking conformity within fixed halo bins at all redshifts,
we find that conformity strength is fairly constant, with low-mass
haloes always showing very weak conformity and high-mass haloes showing
stronger conformity (Figure~ \ref{fig_rad_3}).

\item Our conformity results are generally qualitatively but not quantitatively
consistent between our fiducial simulation and our $25\hmpc$ volume
with $8\times$ better resolution, in part due to the different performance
of the model at different resolutions.
In fact, at our fiducial resolution, our feedback model
appears to overproduce low-mass quenched satellites
(Figure~\ref{fig_sat_mf}), particularly in quenched haloes.  It is
likely that part of the strong one-halo conformity signal in our
massive haloes arises from such over-quenched satellites.
\end{itemize}

Overall, galaxy conformity appears to be a generic prediction of
models that quench massive galaxies approximately in accord with
observations.  However, the strength, extent, and dependence on
specific property are all likely to depend significantly on the
precise physical model driving the quenching.  In our case, we have
tested a heuristic but observationally-consistent scenario where
the hot gas in massive haloes is kept hot, which yields concordant
central galaxy properties, and showed that it has a substantial
impact on galaxy conformity developing in massive haloes.  This
provides a new approach to testing quenching models by examining
neighbouring galaxies, as opposed to tests that focus on the properties
of the central galaxies or their surrounding gas.  It remains to
be seen what the impact of a more physically-motivated quenching
prescription using energy released from AGN would have on conformity;
this will be explored in future work.  This paper presents a first
step towards quantifying conformity in a way that will allow
observations to be compared to models more stringently, and thereby
potentially provide a new way to constrain AGN feedback in modern
galaxy formation models.

\label{conclusion}

\section*{Acknowledgements}
The authors thank G. Kauffmann for providing us with the observational
data and useful ideas for this work, as well as helpful suggestions
from the referee.
The authors also thank F. Durier and T. Naab for helpful conversations and guidance.
MR and RD acknowledge support from the South African Research Chairs
Initiative and the South African National Research Foundation.
MR acknowledges financial support from
Max-Planck-Instit\"ut f\"ur  Astrophysik.
Support for MR was also provided by the Square Kilometre Array
post-graduate bursary program.
The \muf~simulations were run on the Pumbaa astrophysics
computing cluster hosted at the University of the Western Cape,
which was generously funded by UWC's Office of the Deputy Vice
Chancellor. Additional computing resources are obtained from
the Max Planck Computing \& Data Facility (\url{http://www.mpcdf.mpg.de}).




\bibliographystyle{mnras}
\bibliography{paper_bib} 

\begin{thebibliography}{}
\makeatletter
\relax
\def\mn@urlcharsother{\let\do\@makeother \do\$\do\&\do\#\do\^\do\_\do\%\do\~}
\def\mn@doi{\begingroup\mn@urlcharsother \@ifnextchar [ {\mn@doi@}
  {\mn@doi@[]}}
\def\mn@doi@[#1]#2{\def\@tempa{#1}\ifx\@tempa\@empty \href
  {http://dx.doi.org/#2} {doi:#2}\else \href {http://dx.doi.org/#2} {#1}\fi
  \endgroup}
\def\mn@eprint#1#2{\mn@eprint@#1:#2::\@nil}
\def\mn@eprint@arXiv#1{\href {http://arxiv.org/abs/#1} {{\tt arXiv:#1}}}
\def\mn@eprint@dblp#1{\href {http://dblp.uni-trier.de/rec/bibtex/#1.xml}
  {dblp:#1}}
\def\mn@eprint@#1:#2:#3:#4\@nil{\def\@tempa {#1}\def\@tempb {#2}\def\@tempc
  {#3}\ifx \@tempc \@empty \let \@tempc \@tempb \let \@tempb \@tempa \fi \ifx
  \@tempb \@empty \def\@tempb {arXiv}\fi \@ifundefined
  {mn@eprint@\@tempb}{\@tempb:\@tempc}{\expandafter \expandafter \csname
  mn@eprint@\@tempb\endcsname \expandafter{\@tempc}}}

\bibitem[\protect\citeauthoryear{{Ann}, {Park}  \& {Choi}}{{Ann}
  et~al.}{2008}]{Ann-08}
{Ann} H.~B.,  {Park} C.,   {Choi} Y.-Y.,  2008, \mn@doi [\mnras]
  {10.1111/j.1365-2966.2008.13581.x}, \href
  {http://adsabs.harvard.edu/abs/2008MNRAS.389...86A} {389, 86}

\bibitem[\protect\citeauthoryear{{Baldry}, {Glazebrook}, {Brinkmann},
  {Ivezi{\'c}}, {Lupton}, {Nichol}  \& {Szalay}}{{Baldry}
  et~al.}{2004}]{Baldry-04}
{Baldry} I.~K.,  {Glazebrook} K.,  {Brinkmann} J.,  {Ivezi{\'c}} {\v Z}.,
  {Lupton} R.~H.,  {Nichol} R.~C.,   {Szalay} A.~S.,  2004, \mn@doi [\apj]
  {10.1086/380092}, \href {http://adsabs.harvard.edu/abs/2004ApJ...600..681B}
  {600, 681}

\bibitem[\protect\citeauthoryear{{Baldry} et~al.,}{{Baldry}
  et~al.}{2012}]{Baldry-12}
{Baldry} I.~K.,  et~al., 2012, \mn@doi [\mnras]
  {10.1111/j.1365-2966.2012.20340.x}, \href
  {http://adsabs.harvard.edu/abs/2012MNRAS.421..621B} {421, 621}

\bibitem[\protect\citeauthoryear{{Balogh}, {Baldry}, {Nichol}, {Miller},
  {Bower}  \& {Glazebrook}}{{Balogh} et~al.}{2004}]{Balogh-04}
{Balogh} M.~L.,  {Baldry} I.~K.,  {Nichol} R.,  {Miller} C.,  {Bower} R.,
  {Glazebrook} K.,  2004, \mn@doi [\apjl] {10.1086/426079}, \href
  {http://adsabs.harvard.edu/abs/2004ApJ...615L.101B} {615, L101}

\bibitem[\protect\citeauthoryear{{Berti}, {Coil}, {Behroozi}, {Eisenstein},
  {Bray}, {Cool}  \& {Moustakas}}{{Berti} et~al.}{2016}]{Berti-16}
{Berti} A.~M.,  {Coil} A.~L.,  {Behroozi} P.~S.,  {Eisenstein} D.~J.,  {Bray}
  A.~D.,  {Cool} R.~J.,   {Moustakas} J.,  2016, preprint, \href
  {http://adsabs.harvard.edu/abs/2016arXiv160805084B} {} (\mn@eprint {arXiv}
  {1608.05084})

\bibitem[\protect\citeauthoryear{{Bigelow} \& {Dressler}}{{Bigelow} \&
  {Dressler}}{2003}]{Bigelow-03}
{Bigelow} B.~C.,  {Dressler} A.~M.,  2003, in {Iye} M.,  {Moorwood} A.~F.~M.,
  eds,  \procspie Vol. 4841, Instrument Design and Performance for
  Optical/Infrared Ground-based Telescopes. pp 1727--1738,
  \mn@doi{10.1117/12.461870}

\bibitem[\protect\citeauthoryear{{Birnboim} \& {Dekel}}{{Birnboim} \&
  {Dekel}}{2003}]{Birnboim-Dekel-03}
{Birnboim} Y.,  {Dekel} A.,  2003, \mn@doi [\mnras]
  {10.1046/j.1365-8711.2003.06955.x}, \href
  {http://adsabs.harvard.edu/abs/2003MNRAS.345..349B} {345, 349}

\bibitem[\protect\citeauthoryear{{Bray} et~al.,}{{Bray} et~al.}{2016}]{Bray-16}
{Bray} A.~D.,  et~al., 2016, \mn@doi [\mnras] {10.1093/mnras/stv2316}, \href
  {http://adsabs.harvard.edu/abs/2016MNRAS.455..185B} {455, 185}

\bibitem[\protect\citeauthoryear{{Conroy} \& {Gunn}}{{Conroy} \&
  {Gunn}}{2010}]{Conroy-10}
{Conroy} C.,  {Gunn} J.~E.,  2010, {FSPS: Flexible Stellar Population
  Synthesis}, Astrophysics Source Code Library (\mn@eprint {ascl} {1010.043})

\bibitem[\protect\citeauthoryear{{Croton} et~al.,}{{Croton}
  et~al.}{2006}]{Croton-06}
{Croton} D.~J.,  et~al., 2006, \mn@doi [\mnras]
  {10.1111/j.1365-2966.2006.09994.x}, \href
  {http://adsabs.harvard.edu/abs/2006MNRAS.367..864C} {367, 864}

\bibitem[\protect\citeauthoryear{{Dav{\'e}}, {Thompson}  \&
  {Hopkins}}{{Dav{\'e}} et~al.}{2016}]{Dave-16}
{Dav{\'e}} R.,  {Thompson} R.,   {Hopkins} P.~F.,  2016, \mn@doi [\mnras]
  {10.1093/mnras/stw1862}, \href
  {http://adsabs.harvard.edu/abs/2016MNRAS.462.3265D} {462, 3265}

\bibitem[\protect\citeauthoryear{{Dav{\'e}}, {Rafieferantsoa}  \&
  {Thompson}}{{Dav{\'e}} et~al.}{2017a}]{Dave-17b}
{Dav{\'e}} R.,  {Rafieferantsoa} M.~H.,   {Thompson} R.~J.,  2017a, preprint,
  \href {http://adsabs.harvard.edu/abs/2017arXiv170401135D} {} (\mn@eprint
  {arXiv} {1704.01135})

\bibitem[\protect\citeauthoryear{{Dav{\'e}}, {Rafieferantsoa}, {Thompson}  \&
  {Hopkins}}{{Dav{\'e}} et~al.}{2017b}]{Dave-17a}
{Dav{\'e}} R.,  {Rafieferantsoa} M.~H.,  {Thompson} R.~J.,   {Hopkins} P.~F.,
  2017b, \mn@doi [\mnras] {10.1093/mnras/stx108}, \href
  {http://adsabs.harvard.edu/abs/2017MNRAS.467..115D} {467, 115}

\bibitem[\protect\citeauthoryear{{Faucher-Gigu{\`e}re}, {Kere{\v s}},
  {Dijkstra}, {Hernquist}  \& {Zaldarriaga}}{{Faucher-Gigu{\`e}re}
  et~al.}{2010}]{Faucher-Giguere-10}
{Faucher-Gigu{\`e}re} C.-A.,  {Kere{\v s}} D.,  {Dijkstra} M.,  {Hernquist} L.,
    {Zaldarriaga} M.,  2010, \mn@doi [\apj] {10.1088/0004-637X/725/1/633},
  \href {http://adsabs.harvard.edu/abs/2010ApJ...725..633F} {725, 633}

\bibitem[\protect\citeauthoryear{{Gabor} \& {Dav{\'e}}}{{Gabor} \&
  {Dav{\'e}}}{2012}]{Gabor-Dave-12}
{Gabor} J.~M.,  {Dav{\'e}} R.,  2012, \mn@doi [\mnras]
  {10.1111/j.1365-2966.2012.21640.x}, \href
  {http://adsabs.harvard.edu/abs/2012MNRAS.427.1816G} {427, 1816}

\bibitem[\protect\citeauthoryear{{Gabor} \& {Dav{\'e}}}{{Gabor} \&
  {Dav{\'e}}}{2015}]{Gabor-Dave-15}
{Gabor} J.~M.,  {Dav{\'e}} R.,  2015, \mn@doi [\mnras] {10.1093/mnras/stu2399},
  \href {http://adsabs.harvard.edu/abs/2015MNRAS.447..374G} {447, 374}

\bibitem[\protect\citeauthoryear{{Gabor}, {Dav{\'e}}, {Finlator}  \&
  {Oppenheimer}}{{Gabor} et~al.}{2010}]{Gabor-10}
{Gabor} J.~M.,  {Dav{\'e}} R.,  {Finlator} K.,   {Oppenheimer} B.~D.,  2010,
  \mn@doi [\mnras] {10.1111/j.1365-2966.2010.16961.x}, \href
  {http://adsabs.harvard.edu/abs/2010MNRAS.407..749G} {407, 749}

\bibitem[\protect\citeauthoryear{{Geha}, {Blanton}, {Yan}  \& {Tinker}}{{Geha}
  et~al.}{2012}]{Geha-12}
{Geha} M.,  {Blanton} M.~R.,  {Yan} R.,   {Tinker} J.~L.,  2012, \mn@doi [\apj]
  {10.1088/0004-637X/757/1/85}, \href
  {http://adsabs.harvard.edu/abs/2012ApJ...757...85G} {757, 85}

\bibitem[\protect\citeauthoryear{{Hahn} \& {Abel}}{{Hahn} \&
  {Abel}}{2011}]{Hahn-11}
{Hahn} O.,  {Abel} T.,  2011, \mn@doi [\mnras]
  {10.1111/j.1365-2966.2011.18820.x}, \href
  {http://adsabs.harvard.edu/abs/2011MNRAS.415.2101H} {415, 2101}

\bibitem[\protect\citeauthoryear{{Hartley}, {Conselice}, {Mortlock}, {Foucaud}
  \& {Simpson}}{{Hartley} et~al.}{2015}]{Hartley-15}
{Hartley} W.~G.,  {Conselice} C.~J.,  {Mortlock} A.,  {Foucaud} S.,   {Simpson}
  C.,  2015, \mn@doi [\mnras] {10.1093/mnras/stv972}, \href
  {http://adsabs.harvard.edu/abs/2015MNRAS.451.1613H} {451, 1613}

\bibitem[\protect\citeauthoryear{{Hatfield} \& {Jarvis}}{{Hatfield} \&
  {Jarvis}}{2016}]{Hatfield-16}
{Hatfield} P.~W.,  {Jarvis} M.~J.,  2016, preprint, \href
  {http://adsabs.harvard.edu/abs/2016arXiv160608989H} {} (\mn@eprint {arXiv}
  {1606.08989})

\bibitem[\protect\citeauthoryear{{Hearin}, {Watson}  \& {van den
  Bosch}}{{Hearin} et~al.}{2015}]{Hearin-15}
{Hearin} A.~P.,  {Watson} D.~F.,   {van den Bosch} F.~C.,  2015, \mn@doi
  [\mnras] {10.1093/mnras/stv1358}, \href
  {http://adsabs.harvard.edu/abs/2015MNRAS.452.1958H} {452, 1958}

\bibitem[\protect\citeauthoryear{{Hearin}, {Behroozi}  \& {van den
  Bosch}}{{Hearin} et~al.}{2016}]{Hearin-16}
{Hearin} A.~P.,  {Behroozi} P.~S.,   {van den Bosch} F.~C.,  2016, \mn@doi
  [\mnras] {10.1093/mnras/stw1462}, \href
  {http://adsabs.harvard.edu/abs/2016MNRAS.461.2135H} {461, 2135}

\bibitem[\protect\citeauthoryear{{Hopkins}}{{Hopkins}}{2015}]{Hopkins-15}
{Hopkins} P.~F.,  2015, \mn@doi [\mnras] {10.1093/mnras/stv195}, \href
  {http://adsabs.harvard.edu/abs/2015MNRAS.450...53H} {450, 53}

\bibitem[\protect\citeauthoryear{{Kauffmann}}{{Kauffmann}}{2015}]{Kauffmann-15}
{Kauffmann} G.,  2015, \mn@doi [\mnras] {10.1093/mnras/stv2113}, \href
  {http://adsabs.harvard.edu/abs/2015MNRAS.454.1840K} {454, 1840}

\bibitem[\protect\citeauthoryear{{Kauffmann} et~al.,}{{Kauffmann}
  et~al.}{2003}]{Kauffmann-03}
{Kauffmann} G.,  et~al., 2003, \mn@doi [\mnras]
  {10.1046/j.1365-8711.2003.06291.x}, \href
  {http://adsabs.harvard.edu/abs/2003MNRAS.341...33K} {341, 33}

\bibitem[\protect\citeauthoryear{{Kauffmann}, {Li}, {Zhang}  \&
  {Weinmann}}{{Kauffmann} et~al.}{2013}]{Kauffmann-13}
{Kauffmann} G.,  {Li} C.,  {Zhang} W.,   {Weinmann} S.,  2013, \mn@doi [\mnras]
  {10.1093/mnras/stt007}, \href
  {http://adsabs.harvard.edu/abs/2013MNRAS.430.1447K} {430, 1447}

\bibitem[\protect\citeauthoryear{{Kawinwanichakij} et~al.,}{{Kawinwanichakij}
  et~al.}{2016}]{Kawinwanichakij-16}
{Kawinwanichakij} L.,  et~al., 2016, \mn@doi [\apj]
  {10.3847/0004-637X/817/1/9}, \href
  {http://adsabs.harvard.edu/abs/2016ApJ...817....9K} {817, 9}

\bibitem[\protect\citeauthoryear{{Kennicutt}}{{Kennicutt}}{1998}]{Kennicutt-98}
{Kennicutt} Jr. R.~C.,  1998, \mn@doi [\apj] {10.1086/305588}, \href
  {http://adsabs.harvard.edu/abs/1998ApJ...498..541K} {498, 541}

\bibitem[\protect\citeauthoryear{{Kere{\v s}}, {Katz}, {Weinberg}  \&
  {Dav{\'e}}}{{Kere{\v s}} et~al.}{2005}]{Keres-05}
{Kere{\v s}} D.,  {Katz} N.,  {Weinberg} D.~H.,   {Dav{\'e}} R.,  2005, \mn@doi
  [\mnras] {10.1111/j.1365-2966.2005.09451.x}, \href
  {http://adsabs.harvard.edu/abs/2005MNRAS.363....2K} {363, 2}

\bibitem[\protect\citeauthoryear{{Klypin}, {Trujillo-Gomez}  \&
  {Primack}}{{Klypin} et~al.}{2011}]{Klypin-11}
{Klypin} A.~A.,  {Trujillo-Gomez} S.,   {Primack} J.,  2011, \mn@doi [\apj]
  {10.1088/0004-637X/740/2/102}, \href
  {http://adsabs.harvard.edu/abs/2011ApJ...740..102K} {740, 102}

\bibitem[\protect\citeauthoryear{{Krumholz} \& {Gnedin}}{{Krumholz} \&
  {Gnedin}}{2011}]{Krumholz-Gnedin-11}
{Krumholz} M.~R.,  {Gnedin} N.~Y.,  2011, \mn@doi [\apj]
  {10.1088/0004-637X/729/1/36}, \href
  {http://adsabs.harvard.edu/abs/2011ApJ...729...36K} {729, 36}

\bibitem[\protect\citeauthoryear{{Lara-L{\'o}pez}, {Bongiovanni}, {Cepa},
  {P{\'e}rez Garc{\'{\i}}a}, {S{\'a}nchez-Portal}, {Casta{\~n}eda},
  {Fern{\'a}ndez Lorenzo}  \& {Povi{\'c}}}{{Lara-L{\'o}pez}
  et~al.}{2010}]{Lara-Lopez-10}
{Lara-L{\'o}pez} M.~A.,  {Bongiovanni} A.,  {Cepa} J.,  {P{\'e}rez
  Garc{\'{\i}}a} A.~M.,  {S{\'a}nchez-Portal} M.,  {Casta{\~n}eda} H.~O.,
  {Fern{\'a}ndez Lorenzo} M.,   {Povi{\'c}} M.,  2010, \mn@doi [\aap]
  {10.1051/0004-6361/200913886}, \href
  {http://adsabs.harvard.edu/abs/2010A%26A...519A..31L} {519, A31}

\bibitem[\protect\citeauthoryear{{Lawrence} et~al.,}{{Lawrence}
  et~al.}{2007}]{Lawrence-07}
{Lawrence} A.,  et~al., 2007, \mn@doi [\mnras]
  {10.1111/j.1365-2966.2007.12040.x}, \href
  {http://adsabs.harvard.edu/abs/2007MNRAS.379.1599L} {379, 1599}

\bibitem[\protect\citeauthoryear{{Mannucci}, {Cresci}, {Maiolino}, {Marconi}
  \& {Gnerucci}}{{Mannucci} et~al.}{2010}]{Mannucci-10}
{Mannucci} F.,  {Cresci} G.,  {Maiolino} R.,  {Marconi} A.,   {Gnerucci} A.,
  2010, \mn@doi [\mnras] {10.1111/j.1365-2966.2010.17291.x}, \href
  {http://adsabs.harvard.edu/abs/2010MNRAS.408.2115M} {408, 2115}

\bibitem[\protect\citeauthoryear{{McCracken} et~al.,}{{McCracken}
  et~al.}{2012}]{MacCracken-12}
{McCracken} H.~J.,  et~al., 2012, VizieR Online Data Catalog, \href
  {http://adsabs.harvard.edu/abs/2012yCat..35440156M} {354}

\bibitem[\protect\citeauthoryear{{McNamara} \& {Nulsen}}{{McNamara} \&
  {Nulsen}}{2007}]{McNamara-07}
{McNamara} B.~R.,  {Nulsen} P.~E.~J.,  2007, \mn@doi [\araa]
  {10.1146/annurev.astro.45.051806.110625}, \href
  {http://adsabs.harvard.edu/abs/2007ARA%26A..45..117M} {45, 117}

\bibitem[\protect\citeauthoryear{{Mitra}, {Dav{\'e}}  \& {Finlator}}{{Mitra}
  et~al.}{2015}]{Mitra-15}
{Mitra} S.,  {Dav{\'e}} R.,   {Finlator} K.,  2015, \mn@doi [\mnras]
  {10.1093/mnras/stv1387}, \href
  {http://adsabs.harvard.edu/abs/2015MNRAS.452.1184M} {452, 1184}

\bibitem[\protect\citeauthoryear{{Moster}, {Naab}  \& {White}}{{Moster}
  et~al.}{2013}]{Moster-13}
{Moster} B.~P.,  {Naab} T.,   {White} S.~D.~M.,  2013, \mn@doi [\mnras]
  {10.1093/mnras/sts261}, \href
  {http://adsabs.harvard.edu/abs/2013MNRAS.428.3121M} {428, 3121}

\bibitem[\protect\citeauthoryear{{Muratov}, {Keres}, {Faucher-Giguere},
  {Hopkins}, {Quataert}  \& {Murray}}{{Muratov} et~al.}{2015}]{Muratov-15}
{Muratov} A.~L.,  {Keres} D.,  {Faucher-Giguere} C.-A.,  {Hopkins} P.~F.,
  {Quataert} E.,   {Murray} N.,  2015, ArXiv e-prints:1501.03155, \href
  {http://adsabs.harvard.edu/abs/2015arXiv150103155M} {}

\bibitem[\protect\citeauthoryear{{Oppenheimer}, {Dav{\'e}}, {Kere{\v s}},
  {Fardal}, {Katz}, {Kollmeier}  \& {Weinberg}}{{Oppenheimer}
  et~al.}{2010}]{Oppenheimer-10}
{Oppenheimer} B.~D.,  {Dav{\'e}} R.,  {Kere{\v s}} D.,  {Fardal} M.,  {Katz}
  N.,  {Kollmeier} J.~A.,   {Weinberg} D.~H.,  2010, \mn@doi [\mnras]
  {10.1111/j.1365-2966.2010.16872.x}, \href
  {http://adsabs.harvard.edu/abs/2010MNRAS.406.2325O} {406, 2325}

\bibitem[\protect\citeauthoryear{{Peng}, {Lilly}, {Renzini}  \&
  {Carollo}}{{Peng} et~al.}{2012}]{Peng-12}
{Peng} Y.-j.,  {Lilly} S.~J.,  {Renzini} A.,   {Carollo} M.,  2012, \mn@doi
  [\apj] {10.1088/0004-637X/757/1/4}, \href
  {http://adsabs.harvard.edu/abs/2012ApJ...757....4P} {757, 4}

\bibitem[\protect\citeauthoryear{{Planck} et~al.,}{{Planck} et~al.}{2016}]{-16}
{Planck} et~al., 2016, \mn@doi [\aap] {10.1051/0004-6361/201525830}, \href
  {http://adsabs.harvard.edu/abs/2016A%26A...594A..13P} {594, A13}

\bibitem[\protect\citeauthoryear{{Rahmati} \& {Schaye}}{{Rahmati} \&
  {Schaye}}{2014}]{Rahmati-14}
{Rahmati} A.,  {Schaye} J.,  2014, \mn@doi [\mnras] {10.1093/mnras/stt2235},
  \href {http://adsabs.harvard.edu/abs/2014MNRAS.438..529R} {438, 529}

\bibitem[\protect\citeauthoryear{{Schmidt}}{{Schmidt}}{1959}]{Schmidt-59}
{Schmidt} M.,  1959, \mn@doi [\apj] {10.1086/146614}, \href
  {http://adsabs.harvard.edu/abs/1959ApJ...129..243S} {129, 243}

\bibitem[\protect\citeauthoryear{{Sin}, {Lilly}  \& {Henriques}}{{Sin}
  et~al.}{2017}]{Sin-17}
{Sin} L.~P.~T.,  {Lilly} S.~J.,   {Henriques} B.~M.~B.,  2017, preprint, \href
  {http://adsabs.harvard.edu/abs/2017arXiv170208460S} {} (\mn@eprint {arXiv}
  {1702.08460})

\bibitem[\protect\citeauthoryear{{Somerville} \& {Dav{\'e}}}{{Somerville} \&
  {Dav{\'e}}}{2015}]{Somerville-15r}
{Somerville} R.~S.,  {Dav{\'e}} R.,  2015, \mn@doi [\araa]
  {10.1146/annurev-astro-082812-140951}, \href
  {http://adsabs.harvard.edu/abs/2015ARA%26A..53...51S} {53, 51}

\bibitem[\protect\citeauthoryear{{Springel}}{{Springel}}{2005}]{Springel-05}
{Springel} V.,  2005, \mn@doi [\mnras] {10.1111/j.1365-2966.2005.09655.x},
  \href {http://adsabs.harvard.edu/abs/2005MNRAS.364.1105S} {364, 1105}

\bibitem[\protect\citeauthoryear{{Straatman} et~al.,}{{Straatman}
  et~al.}{2014}]{Straatman-14}
{Straatman} C.~M.~S.,  et~al., 2014, \mn@doi [\apjl]
  {10.1088/2041-8205/783/1/L14}, \href
  {http://adsabs.harvard.edu/abs/2014ApJ...783L..14S} {783, L14}

\bibitem[\protect\citeauthoryear{{Tal}, {Quadri}, {Muzzin}, {Marchesini}  \&
  {Stefanon}}{{Tal} et~al.}{2014}]{Tal-14}
{Tal} T.,  {Quadri} R.~F.,  {Muzzin} A.,  {Marchesini} D.,   {Stefanon} M.,
  2014, preprint, \href {http://adsabs.harvard.edu/abs/2014arXiv1405.4856T} {}
  (\mn@eprint {arXiv} {1405.4856})

\bibitem[\protect\citeauthoryear{{Tinker}, {Hahn}, {Mao}, {Wetzel}  \&
  {Conroy}}{{Tinker} et~al.}{2017}]{Tinker-17}
{Tinker} J.~L.,  {Hahn} C.,  {Mao} Y.-Y.,  {Wetzel} A.~R.,   {Conroy} C.,
  2017, preprint, \href {http://adsabs.harvard.edu/abs/2017arXiv170201121T} {}
  (\mn@eprint {arXiv} {1702.01121})

\bibitem[\protect\citeauthoryear{{Tomczak} et~al.,}{{Tomczak}
  et~al.}{2014}]{Tomczak-14}
{Tomczak} A.~R.,  et~al., 2014, \mn@doi [\apj] {10.1088/0004-637X/783/2/85},
  \href {http://adsabs.harvard.edu/abs/2014ApJ...783...85T} {783, 85}

\bibitem[\protect\citeauthoryear{{Vogelsberger} et~al.,}{{Vogelsberger}
  et~al.}{2014}]{Vogelsberger-14}
{Vogelsberger} M.,  et~al., 2014, ArXiv e-prints:1405.2921, \href
  {http://adsabs.harvard.edu/abs/2014arXiv1405.2921V} {}

\bibitem[\protect\citeauthoryear{{Voit}}{{Voit}}{2005}]{Voit-05}
{Voit} G.~M.,  2005, \mn@doi [Advances in Space Research]
  {10.1016/j.asr.2005.02.042}, \href
  {http://adsabs.harvard.edu/abs/2005AdSpR..36..701V} {36, 701}

\bibitem[\protect\citeauthoryear{{Wang} et~al.,}{{Wang} et~al.}{2015}]{Wang-15}
{Wang} J.,  et~al., 2015, \mn@doi [\mnras] {10.1093/mnras/stv1767}, \href
  {http://adsabs.harvard.edu/abs/2015MNRAS.453.2399W} {453, 2399}

\bibitem[\protect\citeauthoryear{{Weinmann}, {van den Bosch}, {Yang}  \&
  {Mo}}{{Weinmann} et~al.}{2006}]{Weinmann-06-a}
{Weinmann} S.~M.,  {van den Bosch} F.~C.,  {Yang} X.,   {Mo} H.~J.,  2006,
  \mn@doi [\mnras] {10.1111/j.1365-2966.2005.09865.x}, \href
  {http://adsabs.harvard.edu/abs/2006MNRAS.366....2W} {366, 2}

\bibitem[\protect\citeauthoryear{{Wetzel}, {Tollerud}  \& {Weisz}}{{Wetzel}
  et~al.}{2015}]{Wetzel-15}
{Wetzel} A.~R.,  {Tollerud} E.~J.,   {Weisz} D.~R.,  2015, \mn@doi [\apjl]
  {10.1088/2041-8205/808/1/L27}, \href
  {http://adsabs.harvard.edu/abs/2015ApJ...808L..27W} {808, L27}

\bibitem[\protect\citeauthoryear{{White} \& {Rees}}{{White} \&
  {Rees}}{1978}]{White-78}
{White} S.~D.~M.,  {Rees} M.~J.,  1978, \mn@doi [\mnras]
  {10.1093/mnras/183.3.341}, \href
  {http://adsabs.harvard.edu/abs/1978MNRAS.183..341W} {183, 341}

\bibitem[\protect\citeauthoryear{{York} et~al.,}{{York} et~al.}{2000}]{York-00}
{York} D.~G.,  et~al., 2000, \mn@doi [\aj] {10.1086/301513}, \href
  {http://adsabs.harvard.edu/abs/2000AJ....120.1579Y} {120, 1579}

\bibitem[\protect\citeauthoryear{{Zu} \& {Mandelbaum}}{{Zu} \&
  {Mandelbaum}}{2017}]{Zu-Mandelbaum-17}
{Zu} Y.,  {Mandelbaum} R.,  2017, preprint, \href
  {http://adsabs.harvard.edu/abs/2017arXiv170309219Z} {} (\mn@eprint {arXiv}
  {1703.09219})

\makeatother
\end{thebibliography}


\label{lastpage}
\end{document}